\documentclass[aps,prd,showpacs,
twocolumn
,superscriptaddress]{revtex4}
\usepackage{graphicx}
\usepackage{dcolumn}
\usepackage{bm}
\usepackage{color}
\usepackage[normalem]{ulem} 
\usepackage[dvipsnames]{xcolor} 
\usepackage{hyperref}
\hypersetup{
  colorlinks=true,        
  linkcolor=blue,         
  citecolor=cyan,         
}

\usepackage{doi}
\usepackage{amsmath}
\usepackage{amssymb}
\usepackage{commath}

\begin{document}

\title{Can the dynamics of test particles around charged stringy black holes\\
mimic the spin of Kerr black holes?}

\author{Bakhtiyor Narzilloev}
\email{nbakhtiyor18@fudan.edu.cn}
\affiliation{Center for Field Theory and Particle Physics and Department of Physics, Fudan University, 200438 Shanghai, China }

\author{Javlon Rayimbaev}
\email{javlon@astrin.uz}
\affiliation{Ulugh Beg Astronomical Institute, Astronomicheskaya 33, Tashkent 100052, Uzbekistan}
\affiliation{National University of Uzbekistan, Tashkent 100174, Uzbekistan}

\author{Sanjar Shaymatov}
\email{sanjar@astrin.uz}
\affiliation{Ulugh Beg Astronomical Institute, Astronomicheskaya 33, Tashkent 100052, Uzbekistan}
\affiliation{National University of Uzbekistan, Tashkent 100174, Uzbekistan}
\affiliation{Institute for Theoretical Physics and Cosmology, Zheijiang University of Technology, Hangzhou 310023, China}
\affiliation{Tashkent Institute of Irrigation and Agricultural Mechanization Engineers, Kori Niyoziy, 39, Tashkent 100000, Uzbekistan}

\author{Ahmadjon~Abdujabbarov}
\email{ahmadjon@astrin.uz}
\affiliation{Ulugh Beg Astronomical Institute, Astronomicheskaya 33, Tashkent 100052, Uzbekistan}
\affiliation{National University of Uzbekistan, Tashkent 100174, Uzbekistan}
\affiliation{Tashkent Institute of Irrigation and Agricultural Mechanization Engineers, Kori Niyoziy, 39, Tashkent 100000, Uzbekistan}
\affiliation{Shanghai Astronomical Observatory, 80 Nandan Road, Shanghai 200030, P. R. China}

\author{Bobomurat Ahmedov}
\email{ahmedov@astrin.uz}
\affiliation{Ulugh Beg Astronomical Institute, Astronomicheskaya 33, Tashkent 100052, Uzbekistan}
\affiliation{National University of Uzbekistan, Tashkent 100174, Uzbekistan}
\affiliation{Tashkent Institute of Irrigation and Agricultural Mechanization Engineers, Kori Niyoziy, 39, Tashkent 100000, Uzbekistan}

\author{Cosimo Bambi}
\email{bambi@fudan.edu.cn}
\affiliation{Center for Field Theory and Particle Physics and Department of Physics, Fudan University, 200438 Shanghai, China }

\date{\today}

\begin{abstract}
We study the motion of electrically charged particles, magnetic monopoles, and magnetic dipoles around electrically and magnetically charged stringy black holes. From the analysis of the radius of the innermost stable circular orbit (ISCO) of electrically charged particles, we show that the electric charge $Q$ of stringy black holes can mimic well the spin of Kerr black holes; the black hole magnetic charge $Q_m$ can mimic spins up to $a_* \simeq 0.85$ for magnetic dipoles; the magnetic charge parameter $g$ of a magnetic monopole can mimic spins up to $a_* \simeq 0.8$. This is due to the destructive character of the magnetic field and such a result excludes the existence of an appreciable black hole magnetic charge in astrophysical black holes from the observation of rapidly rotating objects with dimensionless spin up to $a_* \simeq 0.99$. We then consider the magnetar SGR (PSR) J1745-2900 as a magnetic dipole orbiting the supermassive black hole Sagittarius A* (Sgr~A*). We show that Sgr~A* may be interpreted as a stringy black hole with magnetic charge  $Q_{\rm m}/M \leq 0.4118$. 
\end{abstract}

\maketitle

\section{Introduction}

General Relativity has been extensively tested in weak gravitational fields~\cite{Will06}, and the interest is now shifting to test the theory in the strong field regime~\cite{Berti15,Bambi17c,Cardoso16a,Yagi16,Krawczynski18}. This is also possible thanks to new observational facilities capable of providing unprecedented high quality data. In some specific cases, we have completely new observations, which were impossible up to a few years ago. The Event Horizon Telescope (EHT) collaboration has recently released the direct image of the supermassive black hole in the elliptic galaxy Messier 87 (M87)~\cite{EHT19a,EHT19b}. The LIGO-Virgo experiment can now detect gravitational waves from the coalescence of black holes and neutron stars in compact binaries~\cite{LIGO16a,LIGO16}. On the other hand, we know that General Relativity has a number of theoretical issues demanding new physics, such as the presence of spacetime singularities in almost all physically relevant solutions, the difficulties to find a UV-complete theory of quantum gravity, and some fundamental problems arising in the black hole evaporation process. Observational tests of General Relativity in the strong field regime have thus the potentiality to show us deviations from the predictions of Einstein's gravity and the path towards a UV-complete theory of quantum gravity.

The well-known Kerr-Sen solution proposed in Ref.~\cite{Sen92}  describes black holes in a low-energy limit of heterotic string theory. The solution is specified by the black hole mass $M$, the black hole rotation parameter $a$ (or, equivalently, the dimensionless black hole spin parameter $a_* = a/M$), and an additional electric charge $Q$ associated  to a $U(1)$  gauge  field. 
Optical properties of Kerr-Sen black holes were studied in \cite{An18,Gyulchev07,Younsi16,Hioki08,Dastan16,Uniyal18}. Radial geodesics around Kerr-Sen black holes were studied in~\cite{Blaga01}. In Ref.~\cite{Li07}, the Parikh-Wilczek method was used to study their Hawking radiation. Other thermodynamical properties of Kerr-Sen black holes were extensively investigated in~\cite{Chen09b,Larranaga11,Khani13}. Instabilities of charged massive scalar fields around Kerr-Sen black holes were analyzed in~\cite{Siahaan15}. Particle collisions near Kerr-Sen Dilaton-Axion black holes were studied in~\cite{Debnath15}. Motion of magnetically charged particles were discussed in~\cite{Gonzalez17}. Stringy effects on the relative time delay in the Kerr-Sen and Kerr-Newmann spacetimes were studied in~\cite{Izmailov20,Nathanail2017MNRAS}. The optical properties of a luminous object orbiting near the horizon of a Kerr-Sen black hole were calculated in~\cite{Guo20a}. The approximate final spin and quasinormal modes of a Kerr-Sen black hole resulting from the merger process were analyzed in~\cite{Siahaan20}. 


The properties of the electromagnetic radiation emitted by material orbiting an astrophysical black hole are determined by the motion of massive and massless particles in the strong gravity region of the compact object and the study of these properties can thus be a useful tool to test General Relativity in the strong field regime~\cite{Bambi17c,Krawczynski18}. In General Relativity, the no-hair theorem guarantees that black holes are only characterized by three parameters (mass, spin, and electric charge) and therefore cannot have their own intrinsic magnetic field~\cite{Misner73}. However, a black hole can be embedded into an external magnetic field, and the simplest scenario is that of an external, asymptotically uniform, magnetic field~\cite{Wald74}. In the case of a rotating black hole, the effect of frame dragging alters the structure of the asymptotically uniform magnetic field and an additional electric field appears. The trajectories of charged particles may have a chaotic behavior~ \cite{Chen16,Hashimoto17,Dalui19,Han08,Moura00,MorozovaV2014PhRvD}.
In Refs.~\cite{Jawad16,Hussain15,Jamil15,Hussain17,Babar16,Banados09,Majeed17,Zakria15,Brevik19,DeLaurentis2018PhRvD}, the authors studied the motion of particles in black hole spacetimes in the presence of external magnetic fields. 
The study of the motion of magnetic dipoles in black hole spacetimes in the presence of magnetic fields can be seen as a tool to study the spacetime structure of the strong gravity region around the compact object. The first pioneering attempts to analyze the motion of magnetic dipoles around Schwarzschild and Kerr black holes embedded in external magnetic fields were presented in Refs.~\cite{deFelice,defelice2004}. 
The motion of magnetic dipoles around deformed Schwarzschild black holes in the presence of magnetic fields was studied in~\cite{Rayimbaev16}. Magnetic dipole collisions near rotating black holes in the presence of quintessence were investigated in~\cite{Oteev16}. Acceleration of magnetic dipoles near compact objects in the presence of external magnetic field were explored, for instance, in Refs.~\cite{Toshmatov15d,Abdujabbarov14,Rahimov11a,Rahimov11} in different modified gravity models. 
In our recent papers, we studied the magnetic dipole motion in conformal gravity and in modified gravity models~\cite{Haydarov20,Haydarov20b}.
Studies of the electromagnetic field around black holes in the presence of an external, asymptotically uniform, and dipolar magnetic field are reported in Refs.~\cite{Kolos17,Kovar10,Kovar14,Aliev89,Aliev02,Aliev86,Frolov11,Frolov12,Stuchlik14a,Shaymatov14,Abdujabbarov10,Abdujabbarov11a,Abdujabbarov11,Abdujabbarov08,Karas12a,Shaymatov15,Stuchlik16,Rayimbaev20,Turimov18b,Shaymatov18,Turimov17,Shaymatov20egb,Rayimbaev15,shaymatov19b,Rayimbaev19,Shaymatov20b,Narzilloev2020C,Narzilloev19}.

In the present paper, we study the motion of electrically charged particle, magnetic monopoles, and magnetic dipoles in the strong gravity region of an electrically and magnetically charged Kerr-Sen black hole. The manuscript is organized as follows. 
In Section~\ref{section2}, we study the motion of electrically charged particles around electrically charged stringy black holes. Section~\ref{Sec:metic} is devoted to the dynamics of magnetic monopoles around magnetically charged stringy black holes. We explore the motion of magnetic dipoles around magnetically charged stringy black holes in Section~\ref{chapter1}. Sections~\ref{application} and~\ref{conclusion} are devoted, respectively, to possible astrophysical applications and to our concluding remarks. Throughout the paper, we employ geometrized units in which $G=c=1$ and the spacetime signature $(-,+,+,+)$. Greek (Latin) indices run from 0 to 3 (from 1 to 3).

 \section{Charged particle motion around electrically charged  stringy black holes}\label{section2}

In this section, we study the motion of charged particles around electrically charged, static, stringy black holes, whose spacetime is described by the line element~\cite{Sen92}
\begin{eqnarray}
ds^2&=&-N(r) dt^2+\frac{1}{N(r)} dr^2 +r^2 \left(1+\frac{2 b}{r}\right) d\theta^2 \nonumber\\
&&+r^2 \left(1+\frac{2 b}{r}\right) \sin\theta d\phi^2, \ \\N(r)&=&\left[1-\frac{2 (M-b)}{r}\right]\left(1+\frac{2 b}{r}\right)^{-1}\ ,\nonumber
\end{eqnarray}
where $b=Q^2/2 M$, $Q$ is the black hole electric charge, and $M$ is the black hole mass. One can easily find the radius of the event horizon from $N(r)=0$ and the result is
\begin{eqnarray}
r_h=2 (M-b)\ .
\end{eqnarray}
The event horizon disappears when  $b_{ext}=M$, or, equivalently, when $Q_{ext}=\sqrt{2} M$.
The 4-potential of the electric field around the black hole is
\begin{eqnarray}
A_\mu=\left\{-\frac{Q}{r}\left(1+\frac{2 b}{r}\right)^{-1},\  0, \ 0,\  0\right\}\ .
\end{eqnarray}

The components of the electric field measured by an observer with 4-velocity $u^\alpha$ are given by
\begin{eqnarray}
E_\alpha=F_{\alpha \beta} u^\beta\ ,
\end{eqnarray}
where $F_{\alpha \beta}=A_{\beta;\alpha}-A_{\alpha;\beta}$ is the Faraday tensor and a semicolon denotes a covariant derivate. 
The 4-velocity of the proper observer is 
\begin{eqnarray}
u^\alpha=\left(\sqrt{1+\frac{2 M^2}{M (r-2 M)+Q^2}},0,0,0\right)\ ,
\end{eqnarray}
and the orthonormal components of the electric field are
\begin{eqnarray}\label{eq-ssssss}
E^{\hat{r}}=\frac{M^2 Q }{\left(M
	r+Q^2\right)^2}\ ,\
E^{\hat{\theta}}=E^{\hat{\phi}}=0\ .
\end{eqnarray}
Eq.~(\ref{eq-ssssss}) shows that the only non-vanishing component of the electric field is indeed the radial one, as assumed. One can easily check that in the Newtonian limit (${M}/{r}\rightarrow0$) the above expression reduces to
\begin{eqnarray}
 E^{\hat{r}}=\frac{Q}{r^2}\ ,
\end{eqnarray}
and we recover the Reissner-Nordstr\"{o}m case in the weak field limit.
 
\begin{figure*}[ht!]
	\centering
	\includegraphics[width=0.95\linewidth]{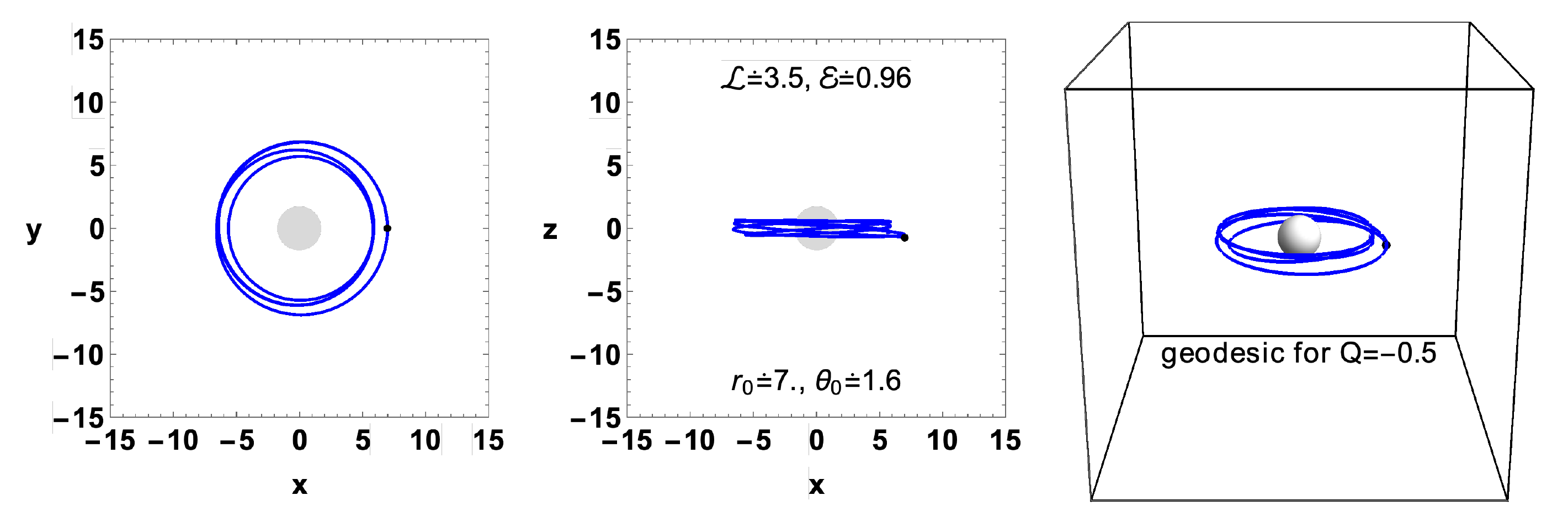}
	\includegraphics[width=0.95\linewidth]{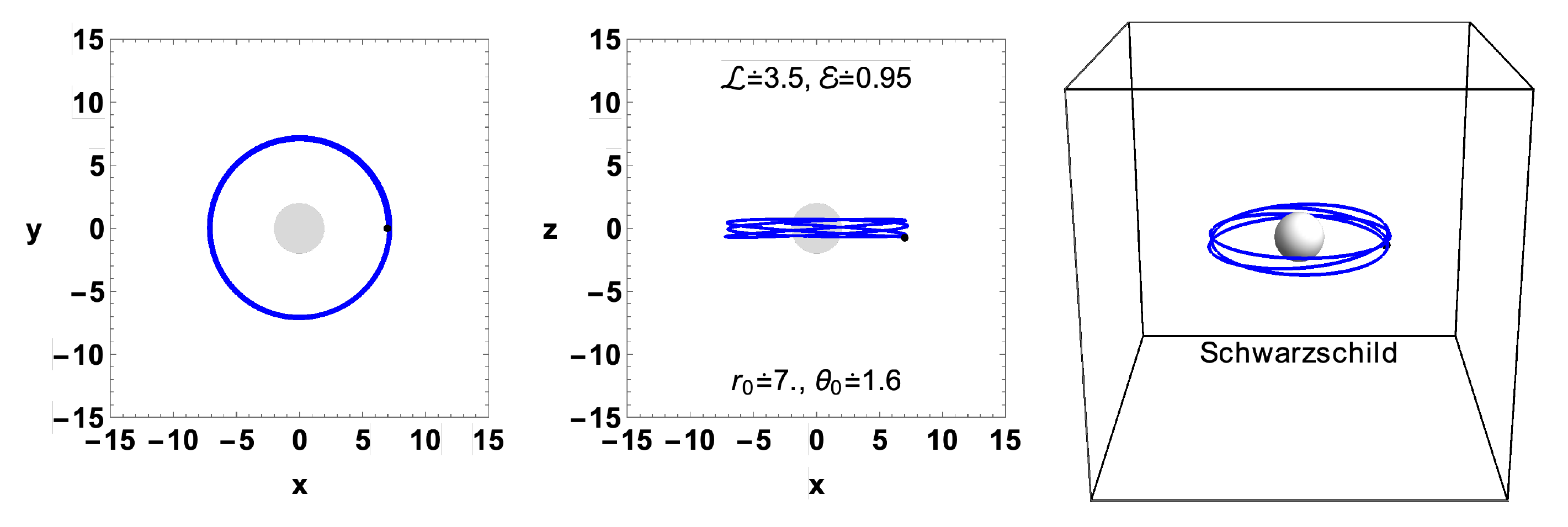}
	\includegraphics[width=0.95\linewidth]{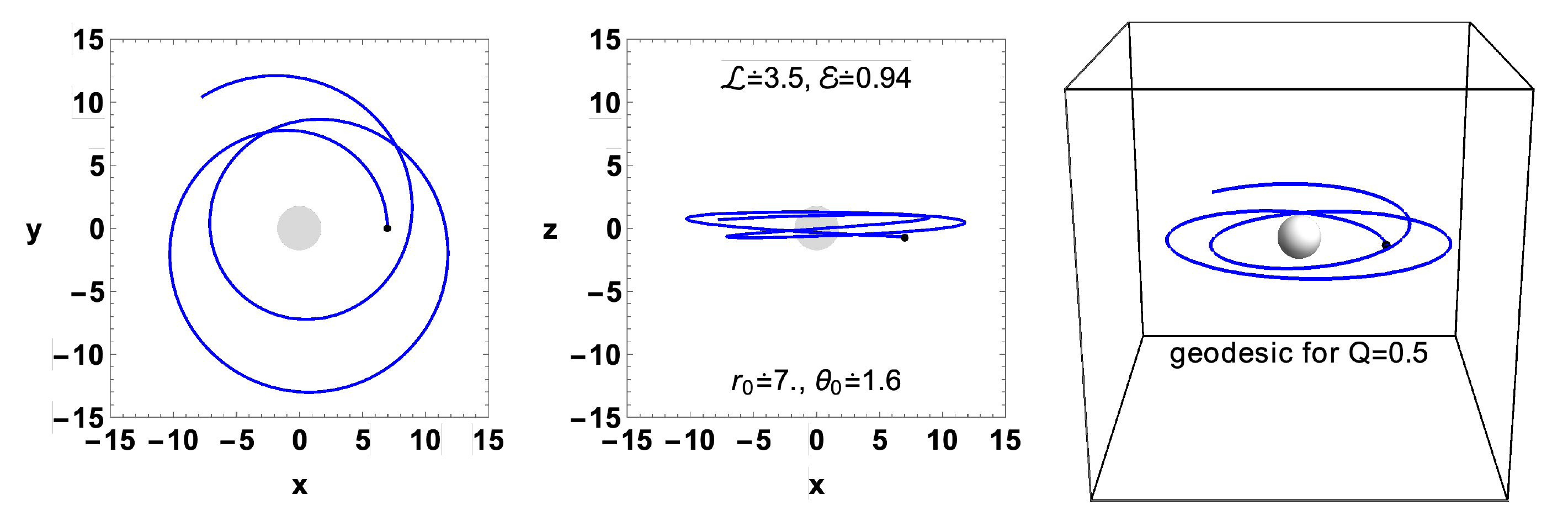}
	\caption{Trajectories of test particles with positive electric charge for different values of the black hole electric charge $Q$. The top row is for a black hole with electric charge $Q/M=-0.5$, the central row is for an electrically neutral Schwarzschild black hole, and the bottom row is for a black hole with electric charge $Q/M=0.5$. }
	\label{trj}
\end{figure*}

Let us now use the Hamilton-Jacobi equation to derive charged particle trajectories. The basic equation is
\begin{eqnarray}\label{Hamjam}
g^{\alpha\beta} \left( \frac{\partial \mathcal{S}}{dx^\alpha}+q A_\alpha\right) \left( \frac{\partial \mathcal{S}}{dx^\beta}+q A_\beta\right)=-m^2\ ,
\end{eqnarray}
where $q$ and $m$ are the electric charge and the mass of the test particle, respectively. 
Considering the symmetries of the spacetime, the action of the test particle $\mathcal{S}$ in (\ref{Hamjam}) can be written as
\begin{eqnarray}
\mathcal{S}=-E t + L \phi + \mathcal{S}_\theta + \mathcal{S}_r\ ,
\end{eqnarray}
where $E$ and $L$ are the energy and angular momentum of the charged particle, respectively. For what follows, it is convenient to introduce the specific energy $\mathcal{E}=E/m$ and specific angular momentum $\mathcal{L}=L/m$. The equation of motion (\ref{Hamjam}) becomes
\begin{eqnarray}
&-&\frac{(2 b \mathcal{E}+\mathcal{E} r+q Q)^2}{N(r) (2 b+r)^2}+N(r) \left(\frac{\partial \mathcal{S}_r}{\partial r}\right)^2\\\nonumber
&+&\frac{\mathcal{L}^2}{(2 b r+r^2) \sin^2\theta} +\frac{1}{2 b r+r^2} \left(\frac{\partial \mathcal{S}_\theta}{\partial \theta}\right)^2=-1\ .
\end{eqnarray}
Fig.~\ref{trj} shows some examples of trajectories for a charged particle. If we increase the value of the electric charge $Q$, we increase the average radius of the trajectory. This fact can be easily interpreted as the result of the interaction between the electric charge of the test particle and the electric charge of the black hole: if both electric charges are positive, the force is repulsive and the average radius increases while, for charges of opposite sign, the force is attractive and the radius decreases.

Now we consider the equatorial plane ($\theta={\pi}/{2}$) as the plane of motion of the charged particle where the effective potential can be defined from the relation 
\begin{eqnarray}\label{V}
\dot{r}^2=\Big[\mathcal{E}-V_{\rm eff}^-(r)\Big]\Big[\mathcal{E}-V_{\rm eff}^{+}(r)\Big]\, ,
\end{eqnarray}
where
\begin{eqnarray}\label{effpotnbm}
&&V_{\rm eff}^{\pm}=\frac{q
	Q}{r} \left(1+\frac{Q^2}{M}\right)^{-1} \\\nonumber
	&&\pm \sqrt{\frac{1-\frac{2M}{r}\left(1-\frac{Q^2}{2 M^2}\right)
		}{1+\frac{Q^2}{M r}}\left[1+\frac{\mathcal{L}^2}{r^2}\left(1+\frac{Q^2}{Mr}\right)^{-1} \right]}\ .
\end{eqnarray}
In what follows, we will only consider $V_{\rm eff}^{+}$, and we will simply write it as $V_{\rm eff}$, because it is the effective potential associated to positive energy orbits.

\begin{figure*}[ht!]
\begin{center}
		\includegraphics[width=0.45\linewidth]{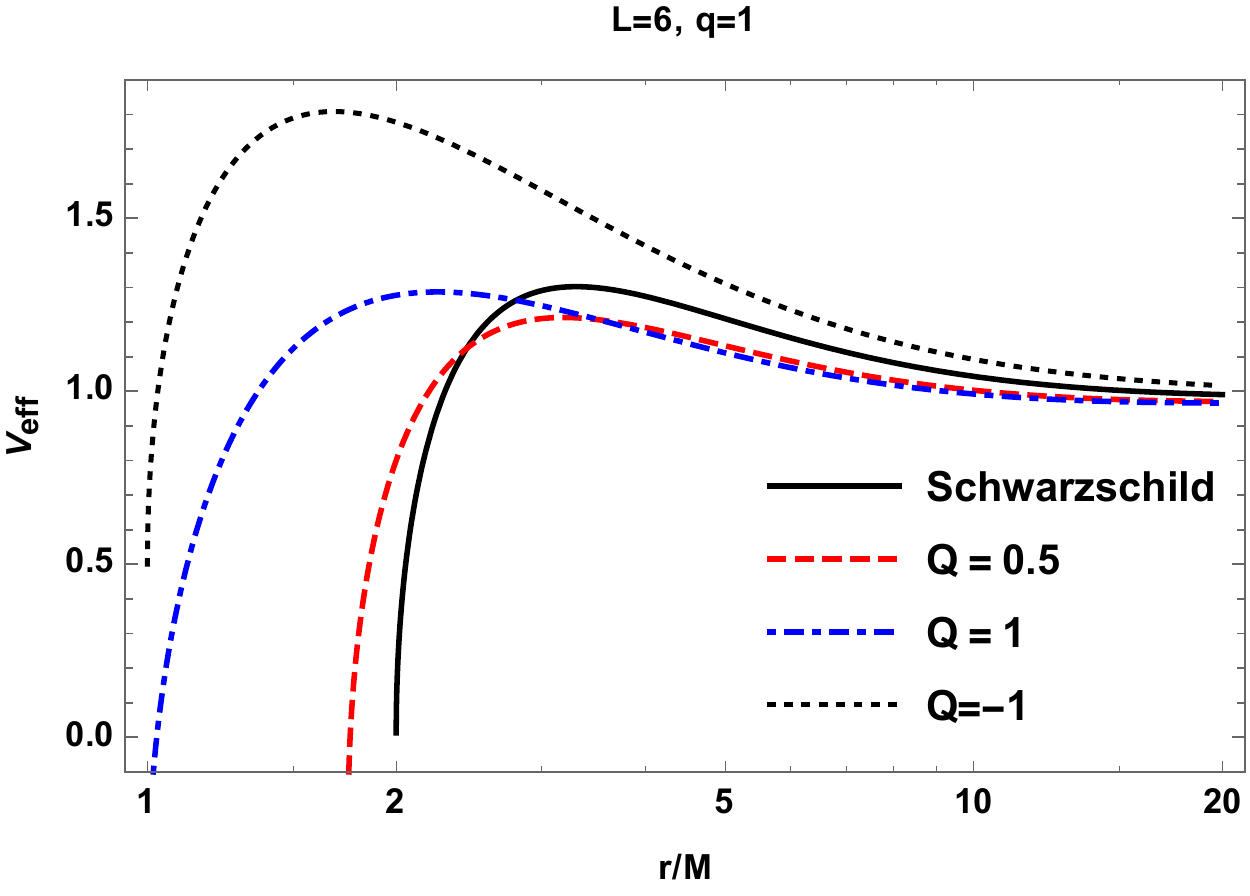}
		\includegraphics[width=0.46\linewidth]{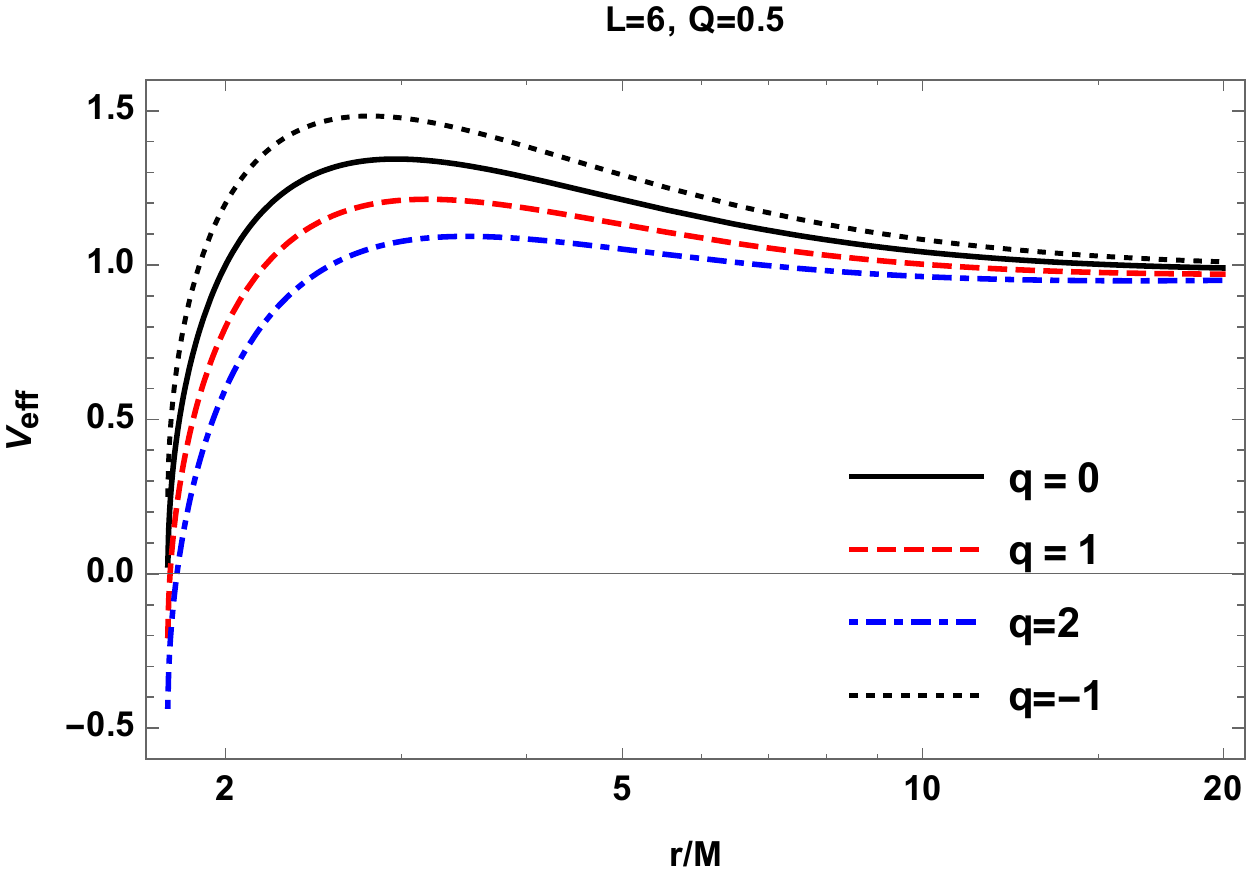}
\end{center}
\caption{Radial profile of the effective potential of charged particles around static and electrically charged stringy black holes in the equatorial plane $\theta=\pi/2$. The left panel is for different values of the black hole charge. The right panel is for different values of the particle charge.  \label{Veff}}
\end{figure*}

The radial profile of the effective potential $V_{\rm eff}$ is shown in Fig.~\ref{Veff}. The electric charges of the stringy black hole and of the test particle can increase or decrease the value of $V_{\rm eff}$ with respect to the Schwarzschild case. Generally speaking, the effective potential decreases if the electric charges of the stringy black hole and of the test particle have the same sign and increases if they have opposite sign. 

\begin{figure*} [ht!]
	\begin{center}
	\includegraphics[width=0.44\linewidth]{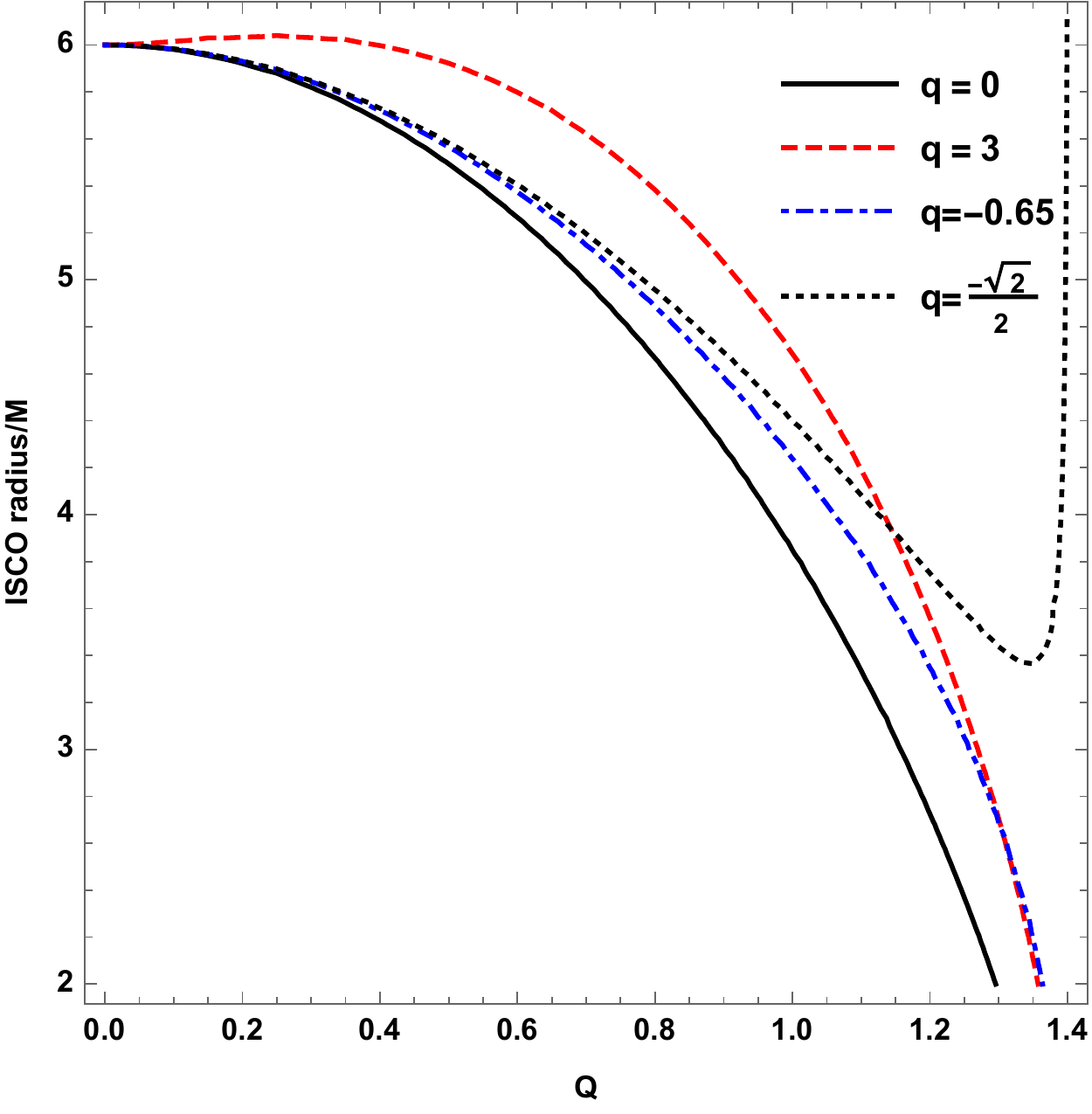}
	\includegraphics[width=0.46\linewidth]{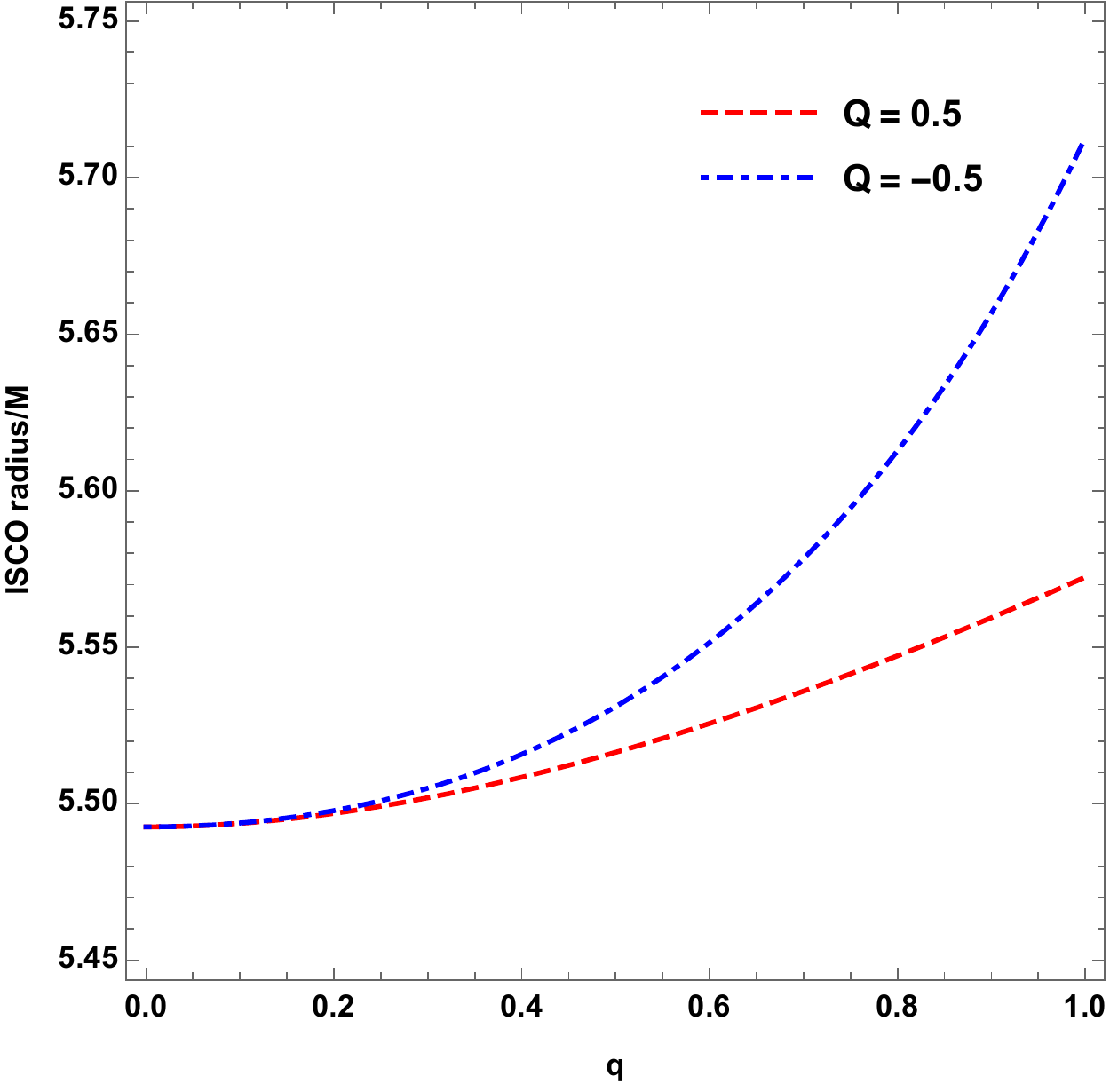}
	\end{center}
	\caption{Left panel: ISCO radius as a function of the black hole electric charge $Q$ for different values of the particle electric charge $q$. Right panel: ISCO radius as a function of the particle electric charge $q$ for different values of the black hole electric charge $Q$.  \label{isco}}
\end{figure*}

Circular orbits are of particular interest because of their astrophysical applications. The Novikov-Thorne model~\cite{Novikov73,Page74} is the standard framework for the description of thin accretion disks around black holes. The model assumes that the particles of the disk follow nearly-geodesic, equatorial, circular orbits, slowly falling into the gravitational well of the black hole. The inner edge of the accretion disk is set at the innermost stable circular orbit (ISCO) and when a particle reaches the ISCO it quickly plunges onto the black hole. The ISCO can thus have important observational implications when it can be associated to the inner edge of the accretion disk of a source. The ISCO radius can be found by solving the following set of equations 
\begin{eqnarray}\label{isco_eqn}
V_{\rm eff}(r)=\mathcal{E}\ ,\ \ 
V_{\rm eff}'(r)=0\ , \ \ 
V_{\rm eff}''(r)=0\ .
\end{eqnarray}
Fig.~\ref{isco} shows the radial coordinate of the ISCO radius as a function of the black hole electric charge $Q$ and the particle electric charge $q$. From the left panel in Fig.~\ref{isco}, we can see quite an intersting phenomenon. If the electric charge of a particle with unit mass is $q=-Q_{\rm ext}/2$, then for a maximally charged black hole ($Q_{\rm ext}=\sqrt{2} M$) the ISCO radius tends to infinity, namely no stable circular orbits can exist, no matter how far the charged particle is from the black hole. The right panel in Fig.~\ref{isco} simply shows that, when the sign of the electric charges of the black hole and the test particle is the same (opposite), the electrostatic interaction decreases (increases) the gravitational force, and therefore the ISCO radius has a smaller (larger) value.

\begin{figure*} [ht!]
	\begin{center}
		\includegraphics[width=0.45\linewidth]{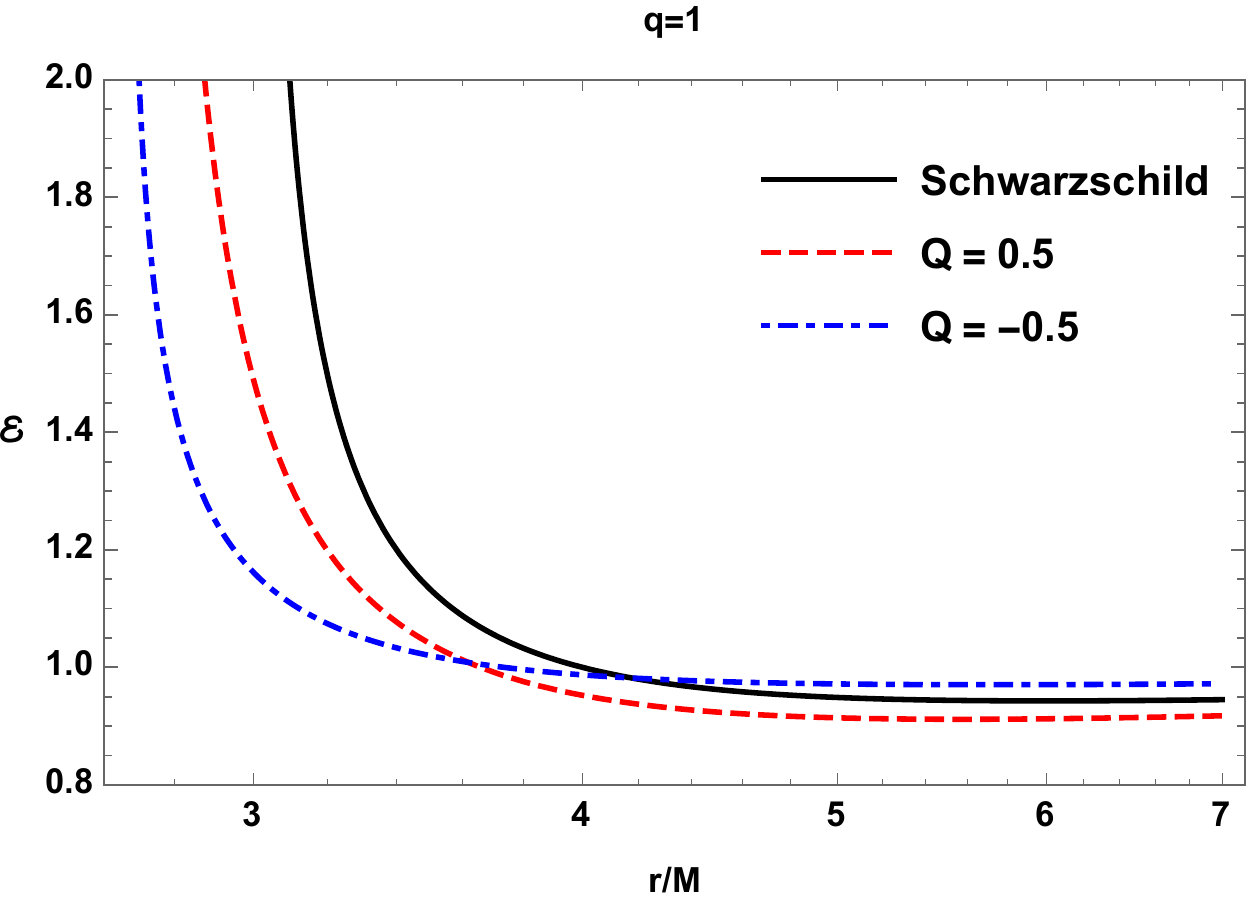}
		\includegraphics[width=0.45\linewidth]{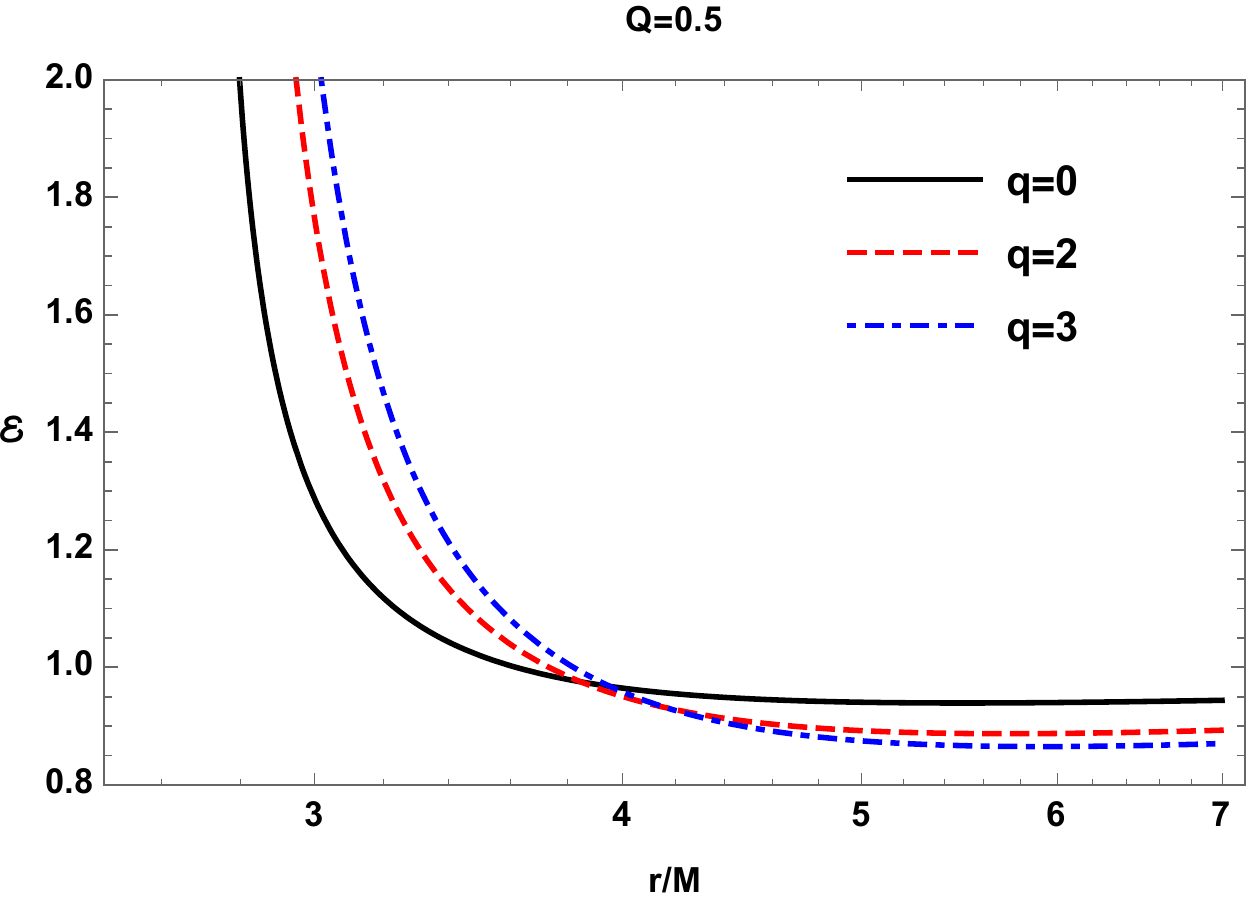}
	\end{center}
	\caption{Energy of an electrically charged particle in a circular orbit as a function of the orbital radius for different values of the black hole electric charge $Q$ (left panel) and of the particle electric charge $q$ (right panel).  \label{E}}
\end{figure*}

\begin{figure*}[ht!]
	\begin{center}
		\includegraphics[width=0.45\linewidth]{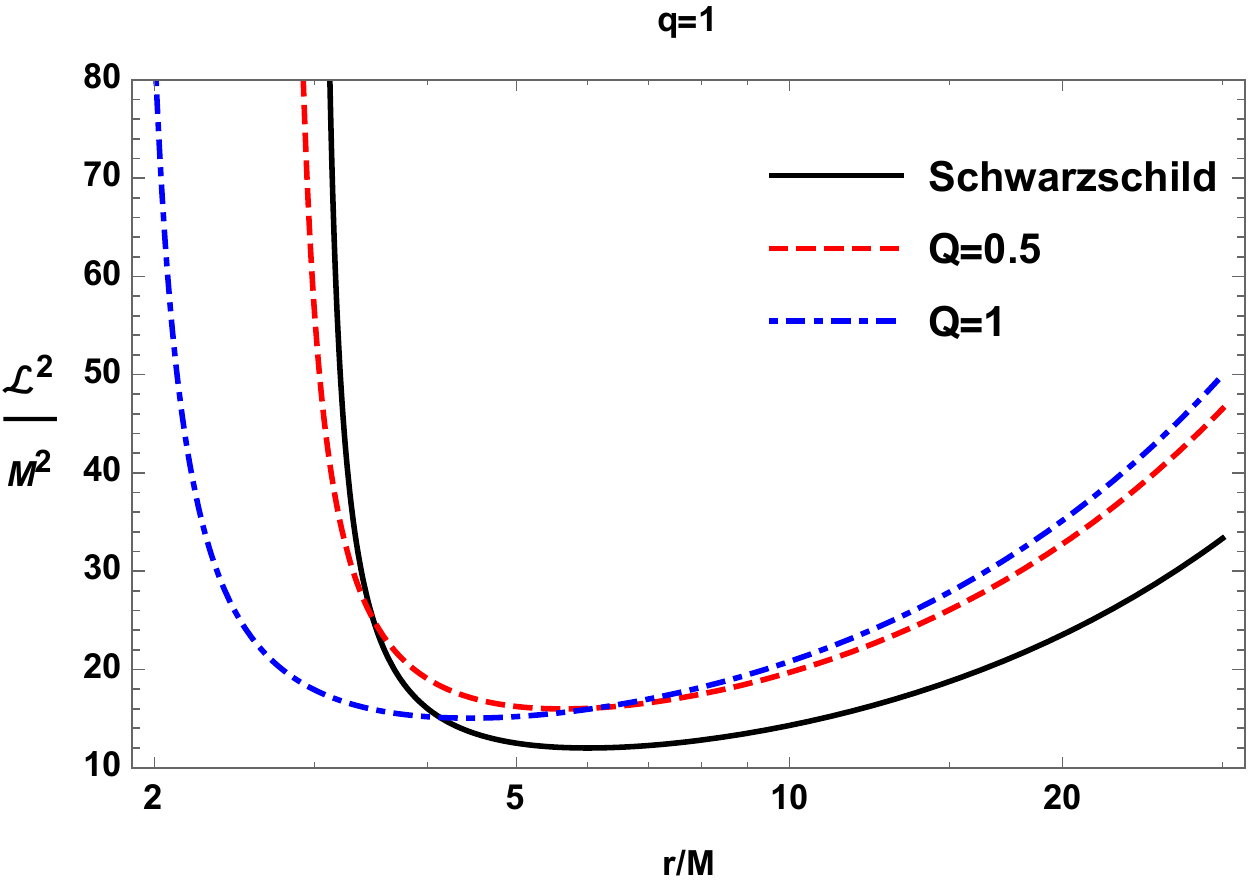}
		\includegraphics[width=0.45\linewidth]{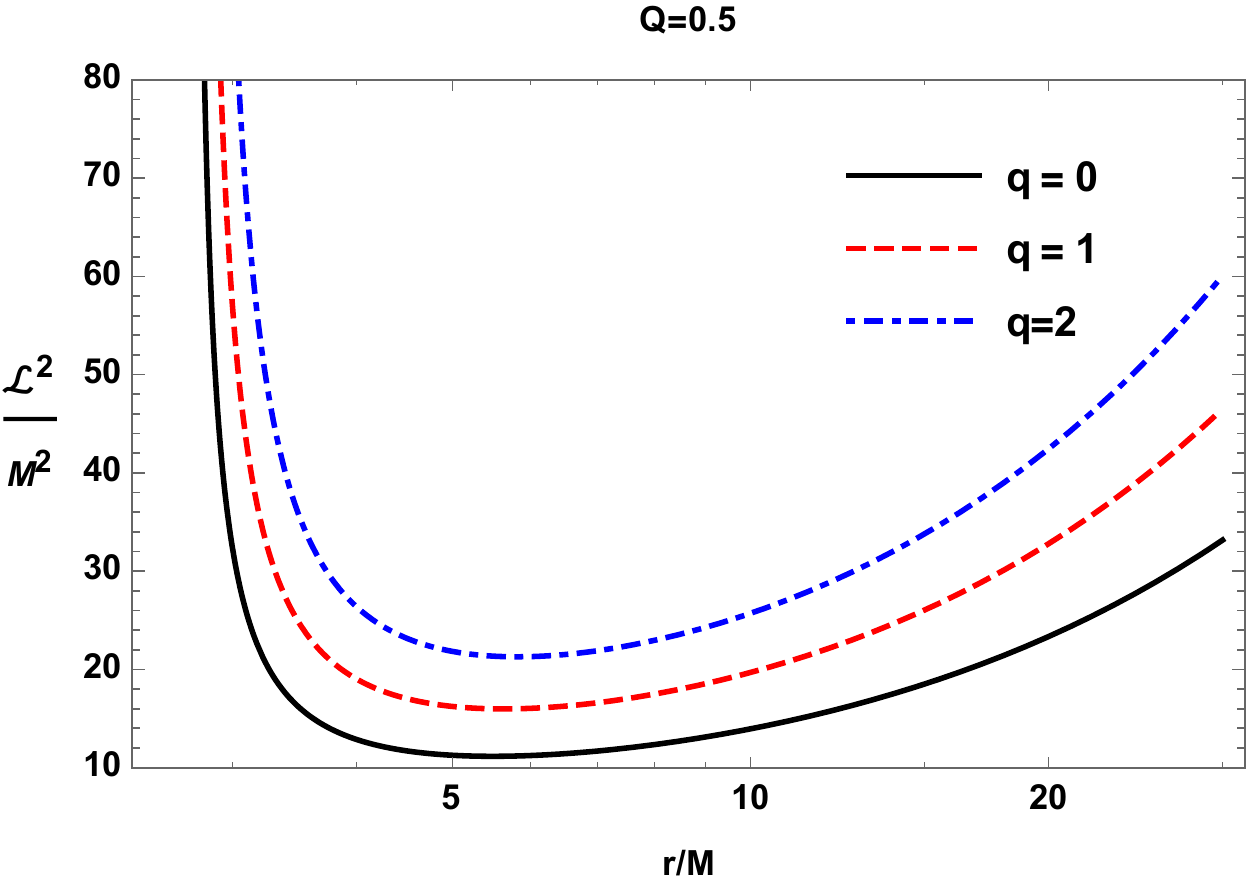}
	\end{center}
	\caption{Angular momentum of an electrically charged particle in a circular orbit as a function of the orbital radius for different values of the black hole electric charge $Q$ (left panel) and of the particle electric charge $q$ (right panel).  \label{L}}
\end{figure*}

The energy of an electrically charge particle in circular orbits is shown in Fig.~\ref{E}. If the attractive force gets stronger, the particle energy decreases. We can also see that the energy of a test particle with opposite charge with respect to the black hole is smaller than the case in which the test particle and the black hole have electric charges of the same sign. The radial profile of the angular momentum of an electrically charged particle in circular orbits is shown in Fig.~\ref{L}. The minimum corresponds to the ISCO radius for that particular parameter configuration.

\begin{figure}[ht!]
\begin{center}
\includegraphics[width=0.95\linewidth]{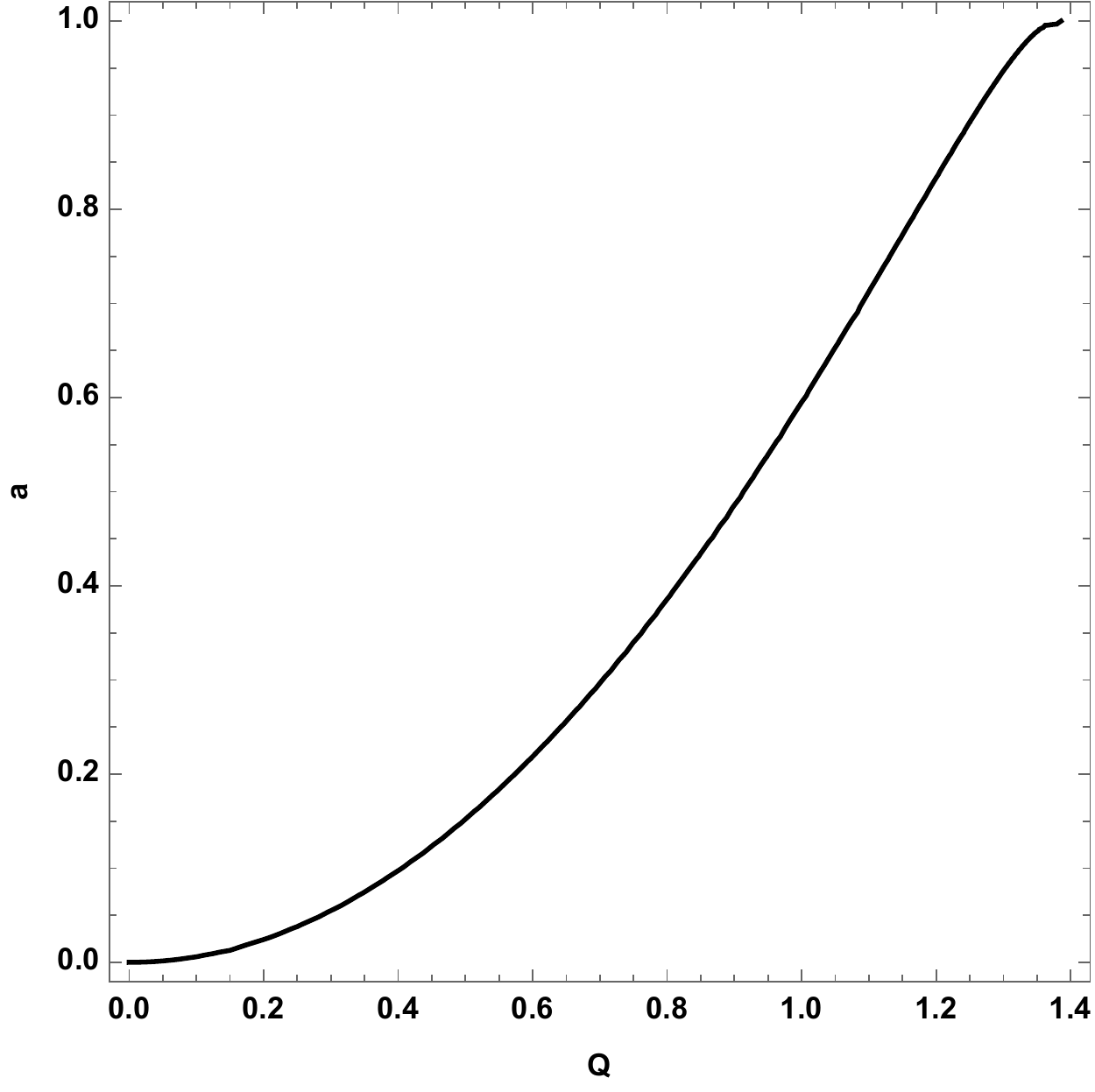}
\end{center}
\caption{Relation between the electric charge $Q$ of a non-rotating stringy black hole and the rotation parameter $a$ of an electrically uncharged Kerr black hole when we require the same value of the ISCO radius (in units of $M$). This plot is obtained in the case of electrically neutral test particles and suggests that a non-rotating stringy black hole can mimic an electrically uncharged Kerr black hole.} \label{aQ}
\end{figure}

It is also remarkable that the electric charge of a static non-rotating black hole can produce similar effects, and thus mimic, the spin parameter of an uncharged Kerr black hole. This is shown in Fig.~\ref{aQ}. The value of the ISCO radius has some important observational effects. However, the same radial coordinate of the ISCO radius can be associated either to a non-rotating but electrically charged stringy black hole and to an electrically uncharged Kerr black hole. It is also remarkable that a maximally charged stringy black hole shares the same ISCO radius with a maximally rotating Kerr black hole; that is, there is a one-to-one correspondence between non-rotating stringy black holes and Kerr black holes. Such a degeneracy suggests that from the observational point of view it may be challenging to distinguish non-rotating stringy black holes from Kerr black holes in the Universe.

We note, however, that astronomical macroscopic objects tends to have a very small electric charge. Because of the highly ionized environment around black holes, we can expect that any possible initial large electric charge can be almost neutralized very quickly to a non-vanishing, but very small, equilibrium electric charge. In such a case, a small electric charge of a stringy black hole cannot mimic a fast-rotating Kerr black hole.

\section{Magnetic monopole motion around magnetically charged stringy black holes
\label{Sec:metic} }

In the previous section, we have discussed the electrically charged stringy black hole metric (also known as Sen black hole metric), which is a solution of heterotic string theory in the four dimensional low-energy field theory limit. However, if the black hole metric is magnetically charged, the spacetime metric is different~\cite{Ong12}.  One can write the action for heterotic string theory in the four dimensional low-energy regime as \cite{Horowitz93,Gonzalez17}
\begin{eqnarray}\label{Eq:action}
\mathcal{S}&=& \int d^4 x \sqrt{-g}\ e^{-2\varphi} \Big[ R+4(\nabla \varphi)^2 -
F_{\alpha\beta}F^{\alpha\beta}  \nonumber\\ &-&\frac{1}{12} H_{\alpha\beta\gamma}
H^{\alpha\beta\gamma}\Big]\, ,  
\end{eqnarray}
where $\varphi$ is the dilaton field and $e^\varphi$ is regarded as a coupling constant reinforcing the stringy effects. Here the 3-form $H_{\alpha\beta\gamma}$ consists of the potential  $B_{\alpha\beta}$ and the Maxwell gauge field $A_\alpha$, which are related by
\begin{eqnarray}
H_{\alpha\beta\gamma}&=&\partial_{\alpha} B_{\beta\gamma}+\partial_{\gamma} B_{\alpha\beta}+\partial_{\beta} B_{\gamma\alpha}\nonumber\\& -& \frac{1}{4}\Big(A_{\alpha} F_{\beta\gamma}+A_{\gamma} F_{\alpha\beta}+A_{\beta} F_{\gamma\alpha}\Big)\, .
\end{eqnarray}

From the action in (\ref{Eq:action}), we can recover the standard Einstein-Hilbert-Maxwell action by setting $H_{\alpha\beta\gamma}=0$. In fact, by rescaling the metric tensor through the coupling constant $g_{\alpha\beta}\rightarrow e^{-2\varphi} g_{\alpha\beta}$, one can rewrite the action in the following form
\begin{eqnarray}\label{Eq:action1}
S=\int d^4x\ \sqrt{-g}\ \left(R - 2(\nabla\varphi)^2 - e^{-2\varphi}
F^2\right)\, . 
\end{eqnarray}
The above action satisfies to the following equation of motion of the Maxwell field
\begin{eqnarray}\label{Eq:eq-motion}
\nabla_\alpha\left(e^{-2\varphi} F^{\alpha\beta}\right) = 0\, . \end{eqnarray}
Note that this equation is invariant under the transformation $F \rightarrow F^{\star}$,
$\varphi\rightarrow -\varphi$. From Eq.~(\ref{Eq:eq-motion}), $F^{\star}_{\alpha\beta} = e^{-2\varphi} \frac{1}{2}
{\epsilon_{\alpha\beta}}^{\gamma\rho}F_{\gamma\rho}$ is satisfied as a curl-free~\cite{Garfinkle91,Gonzalez17}. With such a symmetry, the electromagnetic duality transformation, i.e. $\varphi\rightarrow -\varphi$, can transform an electrically charged black hole solution into a magnetically charged one. Consequently, the spacetime metric describing a magnetically charged black hole in Schwarzschild coordinates $(t, r, \theta, \varphi)$  is written as 
\begin{eqnarray}\label{metric2}
 ds^2= -\frac{f(r)}{h(r)}dt^2+\frac{dr^2}{f(r)h(r)}+r^2d\theta^2+ r^2\sin^2\theta\,d\phi^2,\
\end{eqnarray}
where
\begin{eqnarray}
f(r)=1-\frac{2\, M}{r}\, , \mbox{~~}h(r)=1-\frac{Q^2_m}{M\, r}\, ,
\end{eqnarray}
$M$ is the black hole mass and $Q_m$ is related to the black hole magnetic charge. It is worth noting that the event horizon of the above black hole spacetime is given by 
\begin{eqnarray}
\left(1-\frac{2\, M}{r}\right)\left(1-\frac{Q^2_m}{M\, r}\right)=0\, .
\end{eqnarray}
From the above equation, it is immediately clear that the event horizon is located at $r_{h}=2M$, which is the same radial coordinate of the event horizon as in the Schwarzschild spacetime. We also note that the black hole magnetic charge cannot exceed the value $Q_m=\sqrt{2}  M$.

Let us start with the study of the motion of a magnetic monopole in the background geometry of a magnetically charged stringy black hole. The test particle has rest mass $m$, vanishing electric charge $q=0$, and magnetic charge $q_m$. In general, the Hamiltonian of the system is given by~\cite{Misner73}
\begin{eqnarray}
 H  \equiv  \frac{1}{2}& g^{\alpha\beta}&\left(\frac{\partial \mathcal{S}}{\partial
x^{\alpha}}-q{A}_{\alpha}+iq_m{A}^{\star}_{\alpha}\right)\nonumber\\  &\times &\left(\frac{\partial \mathcal{S}}{\partial
x^{\beta}}-q{A}_{\beta}+iq_m{A}^{\star}_{\beta}\right)\, ,
\label{Eq:H}
\end{eqnarray}
where $\mathcal{S}$ is the action, $x^{\alpha}$ are the spacetime coordinates, $A_{\alpha}$ is the 4-vector potential, and $A^{\star}_{\alpha}$ is the dual 4-vector potential. The non-vanishing components of $A_{\alpha}$ and $A^{\star}_{\alpha}$ are 
\begin{eqnarray}\label{4pots}
A^{\star}_t=-\frac{iQ_m}{r}\,  \mbox{~~~and~~~} A_{\phi}=-Q_m\cos\theta\,  .
\end{eqnarray}
Here we only consider the motion of a magnetic monopole, $q=0$, and thus we are interested in the non-vanishing $A^{\star}_{t}$ of the electromagnetic field as the test particle has only a non-vanishing magnetic charge $q_{m}$~\cite{Gonzalez17}. The case of an electrically charged particle was discussed in the previous section.

The Hamiltonian is a constant and we can set $H=k/2$, where $k$ and $m$ are related by $k=-m^2$. For the above Hamilton-Jacobi equation, the action $\mathcal{S}$ for the motion of a magnetic monopoles in the gravitational field of a magnetically charged black hole has the following form
\begin{eqnarray}\label{Eq:separation1}
\mathcal{S}= -\frac{1}{2}k\lambda-Et+L\varphi+\mathcal{S}_{r}(r)+\mathcal{S}_{\theta}(\theta)\ ,
\end{eqnarray}
where $\mathcal{S}_r$ and $\mathcal{S}_{\theta}$ are functions of only $r$ and $\theta$, respectively. We can then rewrite the Hamilton-Jacobi equation in the following expanded form: 
\begin{eqnarray}\label{Eq:separable}
&-&
\frac{h(r)}{f(r)}\left[-E+
\frac{q_m\,Q_m}{r} \right]^{2}
+f(r)~h(r)
\left(\frac{\partial \mathcal{S}_{r}}{\partial
r}\right)^{2}\nonumber\\&+&{1\over r^{2}}\left(\frac{\partial \mathcal{S}_{\theta}}{\partial
\theta}\right)^{2}+\frac{L^{2}}{r^{2} \sin^2\theta }-k=0 \, . 
\end{eqnarray}
There are four independent constants of motion in the above equation: $E$, $L$, $k$, and a fourth constant related to the latitudinal motion and arising from the separability of the action. However, in this study we ignore the fourth constant of motion because we study the motion of the magnetic monopole on the equatorial plane $\theta=\pi/2$~\cite{Misner73}.
\begin{figure*}
\centering
 \includegraphics[width=0.3\textwidth]{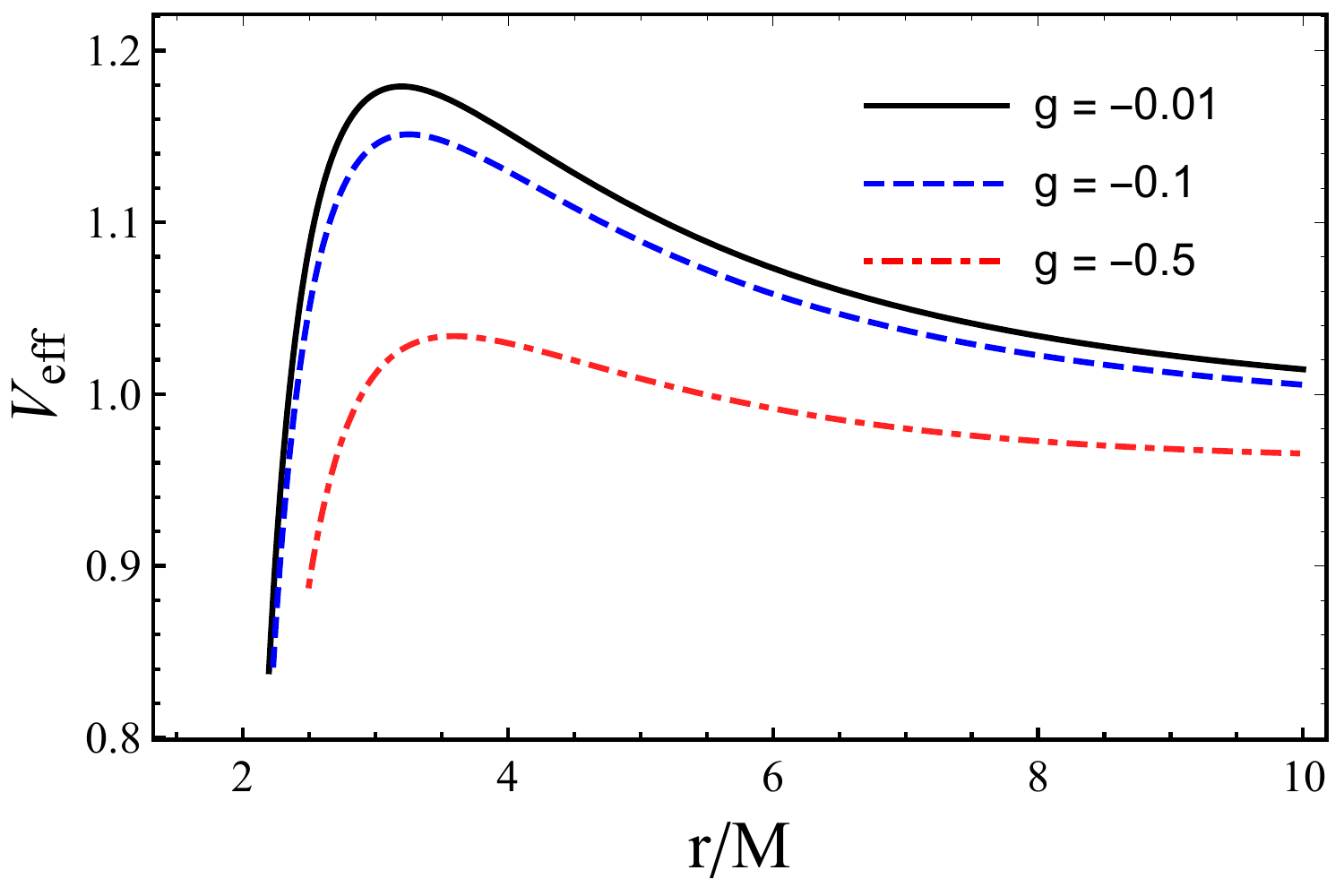}%
 \includegraphics[width=0.3\textwidth]{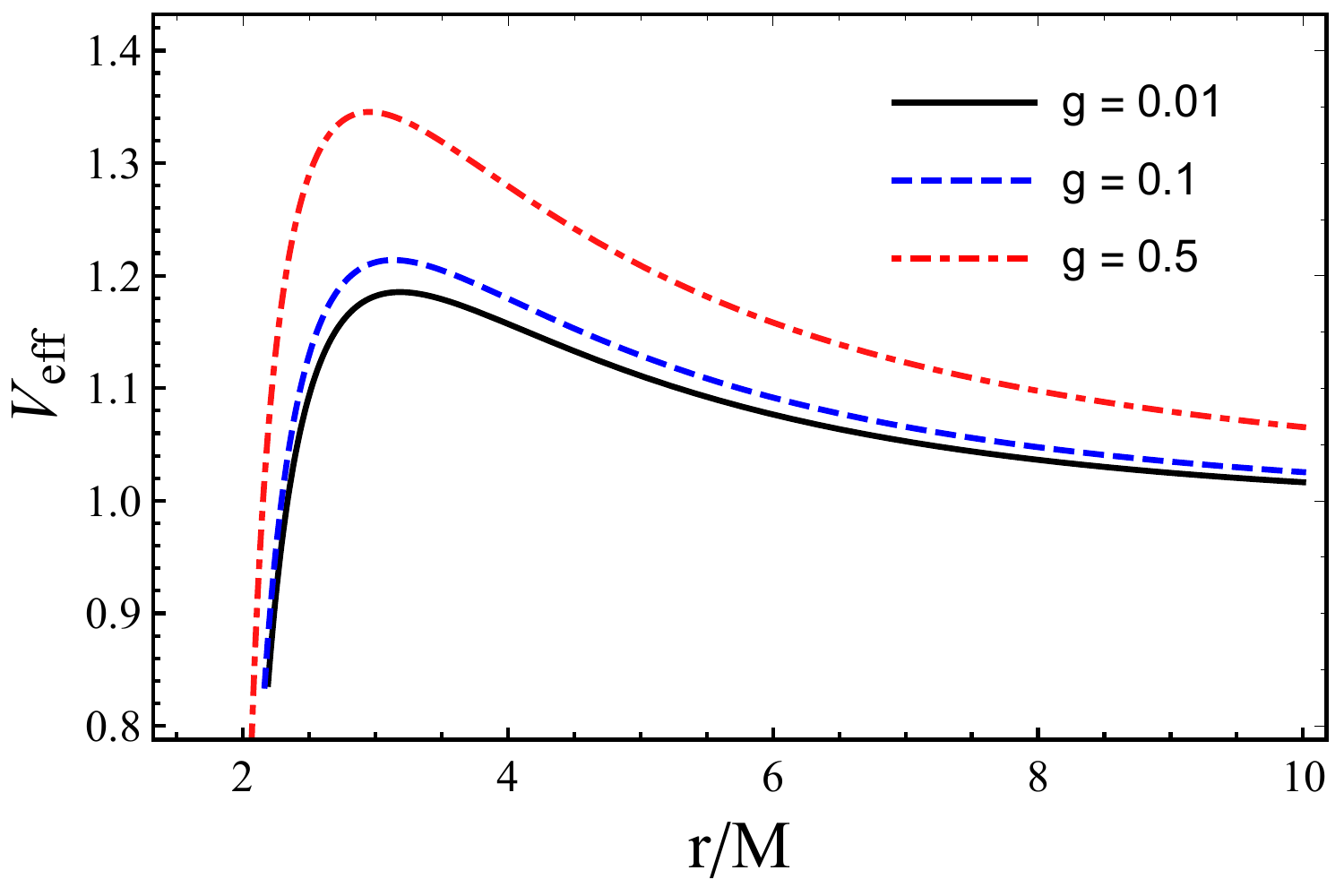}
 \includegraphics[width=0.3\textwidth]{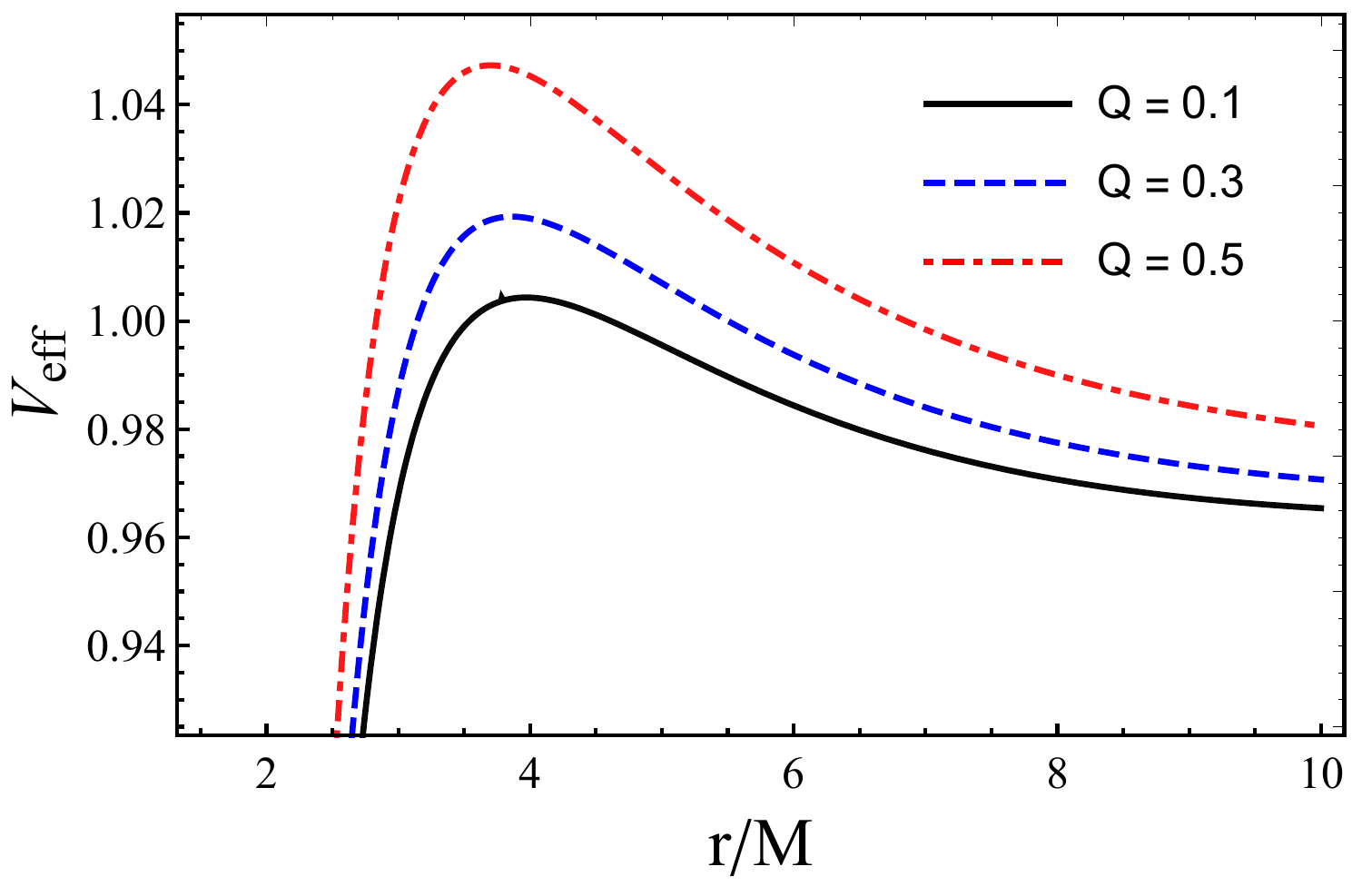}
\caption{\label{fig:eff} Radial profile of the effective potential for massive magnetic monopoles orbiting a magnetically charged black hole. Left and central panels: $V_{eff}$ vs $r/M$ for different values of $g$ in the case of $Q_m=1$. Right panel: $V_{eff}$ vs $r/M$ for different values of $Q_m$ in the case of $g = 0.1$.
}
\end{figure*}
\begin{figure*}

  \includegraphics[width=0.45\textwidth]{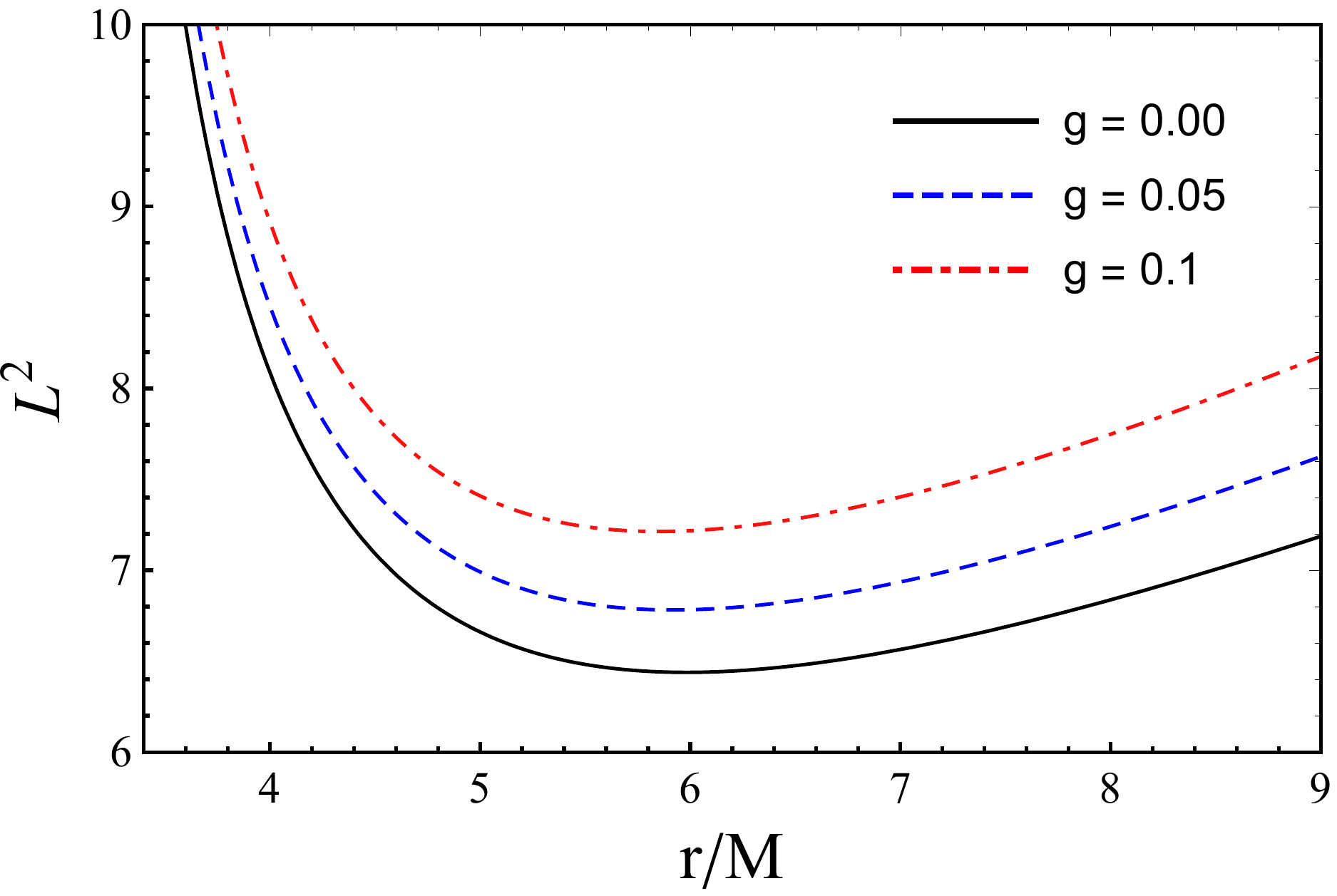}%
  \includegraphics[width=0.45\textwidth]{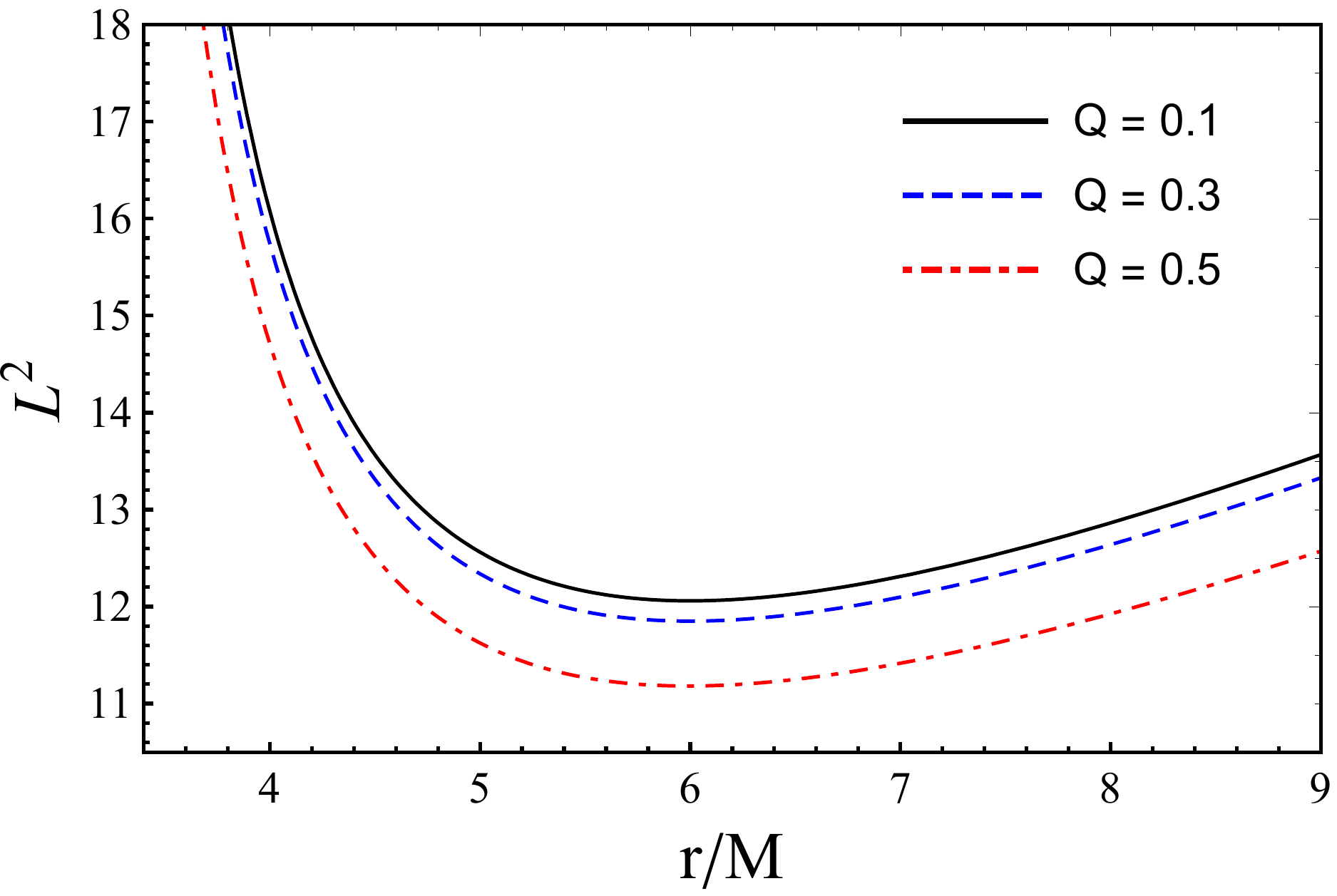}
\caption{\label{fig:ang} Radial profile of the specific angular momentum for magnetic monopoles orbiting a magnetically charged black hole. Left panel: $L^2$ vs $r/M$ for different values of magnetic charge parameter $g$ for $Q_m=1$. Right panel: $L^2$ vs $r/M$ for different values of the black hole magnetic charge $Q_m$ for $g=0.1$. }
\end{figure*}
\begin{table}[ht!]
\caption{\label{T:tab1} Radial coordinates of the ISCO radius for magnetic monopoles orbiting a magnetically charged black hole for different values of the magnetic charge parameter $g$ and black hole magnetic charge $Q_m$. Note that the radial coordinate of the ISCO radius always reduces to the Schwarzschild case, i.e. $r_{isco}=6$, when $g=0$ for any value of $Q_m$. }
\begin{ruledtabular}
\begin{tabular}{c|cccc}
$$  & $$ &  $g$
& $$ & $$  \\ \hline
{$\rm Q_m$}  & $0.01$ &  $0.05$
& $0.10$ & $0.50$  \\
               & $-0.01$ &  $-0.05$ & $-0.1$ & $-0.50$  \\
\hline\\
0.1            &6.00001  &6.00006  &6.00015  &6.00181 \\
               &5.99999  &5.99997  &5.99996  &-  \\\\
0.2            &6.00008 &6.00044 &6.00099 &6.00979  \\
               &5.99992 &5.99967 &5.99944 &-  \\\\
0.5            &6.00136 &6.00719 &6.01545 &6.13338  \\
               &5.99868 &5.99379 &5.98850 &5.97221  \\ \\
0.8            &6.00708 &6.03764 &6.08148 &6.90408  \\
               &5.99313 &5.96765 &5.93985 &5.82584  \\\\
1.0            &6.01869 &6.10156 &6.22713 &7.82272  \\
               &5.98203 &5.91674 &5.84776 &5.57606  \\
\end{tabular}
\end{ruledtabular}
\end{table}
From Eq.~(\ref{Eq:separable}) it is straightforward to derive the radial equation of motion for a magnetic monopole and we find
\begin{eqnarray}\label{Eq:rdot}
\dot{r}^2=\Big[\mathcal{E}-\mathcal{E}_-(r)\Big]\Big[\mathcal{E}-\mathcal{E}_+(r)\Big]\, ,
\end{eqnarray}
where the radial function $\mathcal{E}_{\pm}(r,\mathcal{L},Q_m,g)$
related to the radial motion is given by 
\begin{eqnarray} \label{Veff2}
\mathcal{E}_{\pm}(r,\mathcal{L},Q_m,g)=\frac{gQ_{m}}{r} \pm
\sqrt{\frac{f(r)}{h(r)}\left(1+\frac{\mathcal{L}^2}{r^{2}}\right)}\, ,
\end{eqnarray}
and  we have defined $g=q_m/m$ and $k/m^2=-1$. In Eq.~(\ref{Eq:rdot}), $\dot{r}^2\geq 0$ must always be satisfied, and this implies either $\mathcal{E}>\mathcal{E}_{+}(r,\mathcal{L},Q_m,g)$ or $\mathcal{E}<\mathcal{E}_{-}(r,\mathcal{L},Q_m,g)$. However, we shall restrict ourselves to the positive energy case, which is physically associated to the effective potential, i.e. $V_{\rm eff}(r,\mathcal{L},Q,g)=\mathcal{E}_{+}(r,\mathcal{L},Q_m,g)$. Note that the effective potential reduces to the Schwarzschild one in the case of vanishing magnetic charge $Q_m$. We further focus on the effective potential $V_{\rm eff}(r,\mathcal{L},Q_m,g)$ to determine the motion of the magnetic monopole. 

The radial profile of the effective potential is shown in Fig.~\ref{fig:eff} for different values of $g$ and $Q_m$. As can be seen from the radial profile of  $V_{\rm eff}(r,\mathcal{L},Q_m,g)$, the particle magnetic charge parameter, $g$, and the black hole magnetic charge, $Q_m$, have a similar effects if $g$ is positive. In the case of negative $g$, their effect is opposite. We also note that the strength of the potential for positive $g$ is slightly stronger than the one of the black hole magnetic charge $Q_m$; see Fig.~\ref{fig:eff}.

Let us now consider the circular orbits of a magnetic monopole orbiting a magnetically charged stringy black hole. For circular orbits, we need to consider the following conditions for the effective potential 
\begin{eqnarray}\label{Eq:circular}
V_{\rm eff}(r,\mathcal{L},Q_m,g)=\mathcal{E}, \mbox{~~~}  V_{\rm eff}^{\prime}(r,\mathcal{L},Q_m,g)=0\, ,
\end{eqnarray}
where here a prime $^{\prime}$ denotes a derivative with respect to $r$. It is then straightforward to calculate the specific energy $\mathcal{E}$ and angular momentum $\mathcal{L}$ of a magnetic monopole in a circular orbit. We find
%
\begin{eqnarray}
\mathcal{E}&=&\frac{gQ_{m}}{r} +
\sqrt{\frac{f(r)}{h(r)}\left(1+\frac{\mathcal{L}^2}{r^{2}}\right)}\, , \\
\mathcal{L}^2&=& \frac{r^3 \left[f(r) h'(r)-h(r) f'(r)\right]-2 g^2 Q_m^2 f(r) h(r)^3}{r h(r) f'(r)-f(r) \left[r h'(r)+2 h(r)\right]}\nonumber\\&+&\frac{2 g Q_m f(r) h(r)^2}{\left(r h(r) f'(r)-f(r) \left(r h'(r)+2 h(r)\right)\right)^2}\nonumber\\&\times &
\Big[h(r) \Big(g^2 Q_m^2 h(r)-2 r^3 f'(r)\Big)\nonumber\\&+&2 r^2 f(r) \Big(r h'(r)+2 h(r)\Big)\Big]^{1/2}\, .
\end{eqnarray}
%
The radial profile of the specific angular momentum of a magnetic monopole in a circular orbit around a magnetically charged black hole is shown in Fig.~\ref{fig:ang}. %

Now we determine the ISCO radius of a magnetic monopole orbiting a magnetically charged stringy black hole. We calculate the ISCO radius from the condition $V_{\rm eff}^{\prime\prime}(r,\mathcal{L},Q_m,g)=0$, as done in the previous section. We assume $g=0$. From $V_{\rm eff}^{\prime}(r,\mathcal{L},Q_m,g)=0$, we determine the minimum value of the angular momentum in a circular orbit
\begin{eqnarray}\label{Eq:min-ang}
\mathcal{L}^2=\frac{ \left(Q_m^2-2 M^2\right)r^3 }{6 M^2 r-2 M \left(2 Q_m^2+r^2\right)+Q_m^2 r}\, ,
\end{eqnarray}
\\ 
By substituting Eq.~(\ref{Eq:min-ang}) into $V_{\rm eff}^{\prime\prime}(r,\mathcal{L},Q_m,g)=0$, we find the following condition for the ISCO radius 
\begin{eqnarray}\label{Eq:eq-isco}
\left(Q_m^2-2 M^2 \right) \left(r-6 M \right) \,r^3 \left(Q_m^2 - M r\right)^2=0\, .
\end{eqnarray}
From this equation it is immediately clear that, in the case $g=0$, we have $r_{isco}=6M$, which is the same value as in the Schwarzschild spacetime. The observational implications are straightforward: a faraway observer cannot distinguish a magnetically charged stringy black hole from a Schwarzschild black hole. Table~\ref{T:tab1} shows the numerical results of the ISCO radius for different values of the magnetic charge parameter $g$ and of the black hole magnetic charge $Q_m$. If $g$ is positive, the ISCO radius increases as $g$ increases. If $g$ is negative, the ISCO radius decreases as $|g|$ increases. Note that for $|g|\ll 1$ we can write $r_{\rm isco}=6\pm\delta r_{\rm isco}$, namely for small values of the magnetic charge parameter there are only small changes in the value of the ISCO radius.

\begin{figure}

  \includegraphics[width=0.45\textwidth]{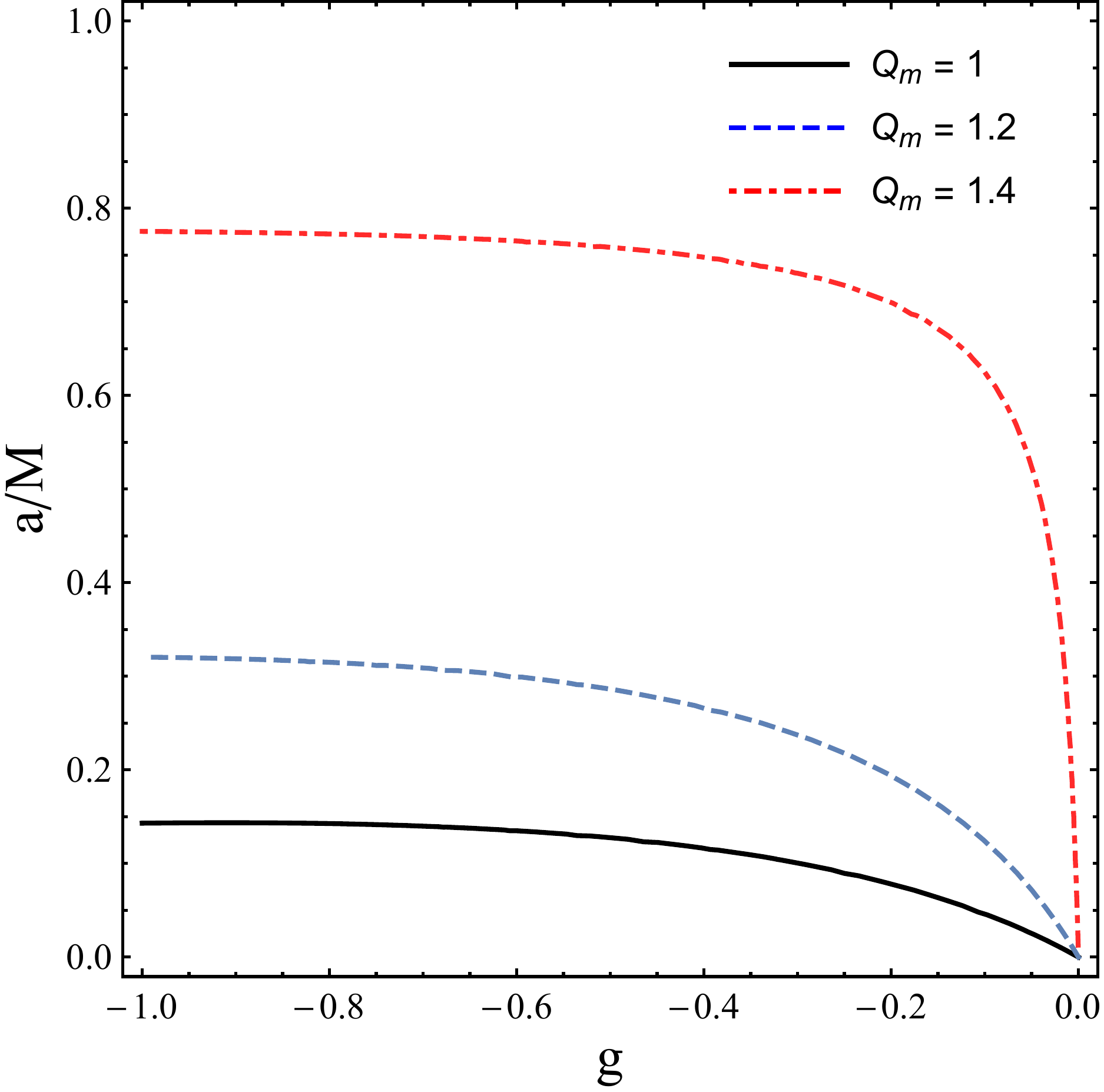}
\caption{Rotation parameter $a$ of a Kerr black hole vs magnetic charge $g$ of a test particle orbiting a stringy black hole with magnetic charge $Q_m$ leading to the same ISCO radius for different values of $Q_m$. However, the stringy scenario can mimic the standard scenario of a Kerr black hole only for a Kerr black hole spin parameter up to 0.8. 
\label{fig:a_vs_g} }
\end{figure}

We now calculate the value of the magnetic charge $g$ of a test particle orbiting around a magnetically charged stringy black hole and the value of the rotation parameter $a$ of a Kerr black hole leading to the same ISCO radius. Our results are shown in Fig.~\ref{fig:a_vs_g}. However, unlike the previous case of an electrically charged particle around an electrically charged stringy black hole, it is not possible to reproduce the ISCO value of any Kerr black hole. It is only possible to recover the ISCO radius of Kerr black holes with spin parameters up to $ a/M \approx 0.8$, but it is impossible to get smaller values of the ISCO radius as in the case of faster rotating Kerr black holes. 

\section{Magnetic dipoles around magnetically charged stringy black holes \label{chapter1}}

In this section, we study the dynamics of magnetic dipoles orbiting magnetically charged stringy black holes with the spacetime metric given in~(\ref{metric2}) and the electromagnetic potential in~(\ref{4pots}). The non-vanishing components of the Faraday tensor  $F_{\mu\nu}$ are
\begin{eqnarray}\label{FFFF}
 F_{\theta \phi}= - F_{\phi\theta} = -Q_m \sin\theta \ .
\end{eqnarray}

The magnetic field of the magnetically charged stringy black hole can be derived from
\begin{eqnarray}\label{fields}
B^{\alpha} &=& \frac{1}{2} \eta^{\alpha \beta \sigma \mu} F_{\beta \sigma} w_{\mu}\ ,
\end{eqnarray}
where $w_{\mu}$ is the 4-velocity of observer, $\eta_{\alpha \beta \sigma \gamma}$ is the pseudo-tensorial form of the Levi-Civita symbol $\epsilon_{\alpha \beta \sigma \gamma}$ defined as
\begin{eqnarray}
\eta_{\alpha \beta \sigma \gamma}=\sqrt{-g}\epsilon_{\alpha \beta \sigma \gamma}\ , \qquad \eta^{\alpha \beta \sigma \gamma}=-\frac{1}{\sqrt{-g}}\epsilon^{\alpha \beta \sigma \gamma}\ ,
\end{eqnarray}
where $g={\rm det|g_{\mu \nu}|}=-r^4\sin^2\theta$ for the spacetime metric in~(\ref{metric2}), $\epsilon_{0123} = 1$, and
\begin{eqnarray}
\epsilon_{\alpha \beta \sigma \gamma}=\begin{cases}
+1\ , \rm for\  even \ permutations\ ,
\\
-1\ , \rm for\  odd\  permutations\ ,
\\
\ \ 0 \ , \rm for\ the\ other\ combinations\ ,
\end{cases}\ .
\end{eqnarray}

The orthonormal radial component of the magnetic field of the magnetically charged stringy black hole is
\begin{equation}\label{BrBt}
    B^{\hat{r}}=\frac{Q_m}{r^2} \ .
\end{equation}
Eq. (\ref{BrBt}) implies that the radial component of the magnetic field around a magnetically charged black hole is not effected by the spacetime geometry of the stringy black hole and formally coincides with the standard Newtonian expression.
    
We can study the dynamics of magnetic dipoles around magnetically charged black holes using the Hamilton-Jacobi equation \cite{deFelice}
\begin{eqnarray}\label{HJ}
g^{\mu \nu}\frac{\partial {\cal S}}{\partial x^{\mu}} \frac{\partial {\cal S}}{\partial x^{\nu}}=-\Bigg(m-\frac{1}{2} {\cal D}^{\mu \nu}F_{\mu \nu}\Bigg)^2\ ,
\end{eqnarray}
where the term ${\cal D}^{\mu \nu}{\cal F}_{\mu \nu}$ is responsible for the interaction between the magnetic dipole and the magnetic field generated by the magnetic charge of the stringy black hole. Here we assume that the magnetic dipole has the corresponding polarization tensor ${\cal D}^{\alpha \beta}$ that satisfies the following conditions
\begin{eqnarray}\label{dexp}
{\cal D}^{\alpha \beta}=\eta^{\alpha \beta \sigma \nu}u_{\sigma}\mu_{\nu}\ , \qquad {\cal D}^{\alpha \beta }u_{\beta}=0\ ,
\end{eqnarray} 
where $\mu^{\nu}$ is the dipole moment of the magnetic dipole. Here we determine the interaction term ${\cal D}^{\mu \nu}{\cal F}_{\mu \nu}$ using the relation between the Faraday tensor $F_{\alpha \beta}$ and the components of the electric field, $E_{\alpha}$, and of the magnetic field, $B^{\alpha}$,
\begin{eqnarray}\label{fexp}
F_{\alpha \beta}=w_{\alpha}E_{\beta}-w_{\beta}E_{\alpha}-\eta_{\alpha \beta \sigma \gamma}w^{\sigma}B^{\gamma}\ .
\end{eqnarray}

Employing the condition given in Eq.~(\ref{dexp}) and the non-zero components of the Faraday tensor, we have 
\begin{eqnarray}\label{DF1}
{\cal D}^{\alpha \beta}F_{\alpha \beta}=2\mu_{\alpha}B^{\alpha}=2{\cal \mu}^{\hat{\alpha}}B_{\hat{\alpha}} \ .
\end{eqnarray}
We assume that the motion of the test particle is on the equatorial plane and that its magnetic dipole moment is aligned along the direction of magnetic field lines of the stringy black hole. In such a case, the components of the dipole magnetic moment of the particle are $\mu^{i}=(\mu^{r},0,0)$. This configuration allows for an equilibrium state for the interaction between the magnetic field and the magnetic dipole, while other configurations cannot provide any stable equilibrium state. This configuration also allows to study the particle motion and  we may avoid the  relative motion problem by choosing the appropriate observer's frame. Due to the constant value of the magnetic moment of the particle, the second condition in~(\ref{dexp}) is automatically satisfied. The interaction part can be calculated using Eqs.~(\ref{DF1}) and (\ref{FFFF})
\begin{eqnarray}\label{DF3} {\cal D}^{\alpha \beta}F_{\alpha \beta} = \frac{2\mu Q_m}{r^2} \ , \end{eqnarray}
where $\mu = \sqrt{\vline \mu_{\hat{i}}\mu^{\hat{i}} \vline }$ is the norm of the magnetic dipole moment of the particle.

Due to the symmetries of the magnetic field and of the spacetime, we can write the action of the magnetic dipole in the Hamilton-Jacobi equation (\ref{HJ}) in the following form
\begin{eqnarray}\label{action}
{\cal S}=-E t+L \phi +{\cal S}_{r\theta}(r,\theta)\ .
\end{eqnarray}
Since we consider the motion on the equatorial plane ($\theta=\pi/2$), Eqs.~(\ref{DF1}), (\ref{HJ}) and (\ref{action}) provide the following equation for the radial component 
\begin{eqnarray}
\dot{r}^2={\cal{E}}^2-V_{\rm eff}(r;{\cal L},{\cal B})\ , 
\end{eqnarray}
where the effective potential has the form
\begin{eqnarray}\label{effpot}
V_{\rm eff}(r;{\cal L},{\cal B})=\frac{f(r)}{h(r)}\left[\left(1-\frac{{\cal B}}{r^2}\right)^2+\frac{{\cal L}^2}{r^2}\right]\,
\end{eqnarray}
where $${\cal B} =\frac{\mu Q_m}{m}\ ,$$ is the magnetic interaction parameter responsible for the interaction between the magnetic dipole of the test particle and the proper magnetic field of the magnetically charged stringy black hole. $\beta=\mu/(m M)$ is a dimensionless parameter characterizing the magnetic dipole and the spacetime parameters.  $\beta$ is always  positive. Modeling a magnetized neutron star as a test magnetic dipole with moment $\mu = (1/2)B_{\rm NS}R^3_{\rm NS}$ orbiting around a supermassive black hole (SMBH), we find
\begin{eqnarray}\nonumber
   \beta&=&\frac{B_{\rm NS}R_{\rm NS}^3}{2m_{\rm NS}M_{\rm SMBH}}= 0.0341 \left(\frac{B_{\rm NS}}{10^{12}\rm G}\right)\left(\frac{R_{\rm NS}}{10^6 \rm cm}\right)\\
   &&\times \left(\frac{m_{\rm NS}}{1.4 M_{\odot}}\right)^{-1}\left(\frac{M_{\rm SMBH}}{3.8\cdot 10^6M_{\odot}}\right)^{-1}\ .
\end{eqnarray}

The circular stable orbits of the magnetic dipole around the central object can be derived by the  conditions
\begin{eqnarray} \label{conditions}
V_{\rm eff}'=0\ , V_{\rm eff}'' \geq 0  \ ,
\end{eqnarray}
which can be used to find the specific angular momentum and the specific energy in circular orbits of the magnetic dipole:
\begin{eqnarray}
\nonumber
{\cal L}^2&=&\frac{\left(r^2-\mathcal{B}\right)}{r^2 \mathcal{F}(r) }\Big\{2 M^2 \left(r^3-5 r \mathcal{B}\right)+4 M \mathcal{B} \left(2 Q^2_{m}+r^2\right)
\\
&-&Q^2_{m} r \left(r^2+3 \mathcal{B}\right)\Big\}\ ,
\\
{\cal E}^2&=&\frac{2 M (r-2 M)^2 }{\mathcal{F}(r)}\left(1-\frac{\mathcal{B}^2}{r^4}\right)\ , 
\end{eqnarray}
where $\mathcal{F}(r)=2 M (Q^2_{m}+r^2)-r \left(6 M^2+Q^2_{m}\right)$. 

\begin{figure}[ht!]\centering\includegraphics[width=0.98\linewidth]{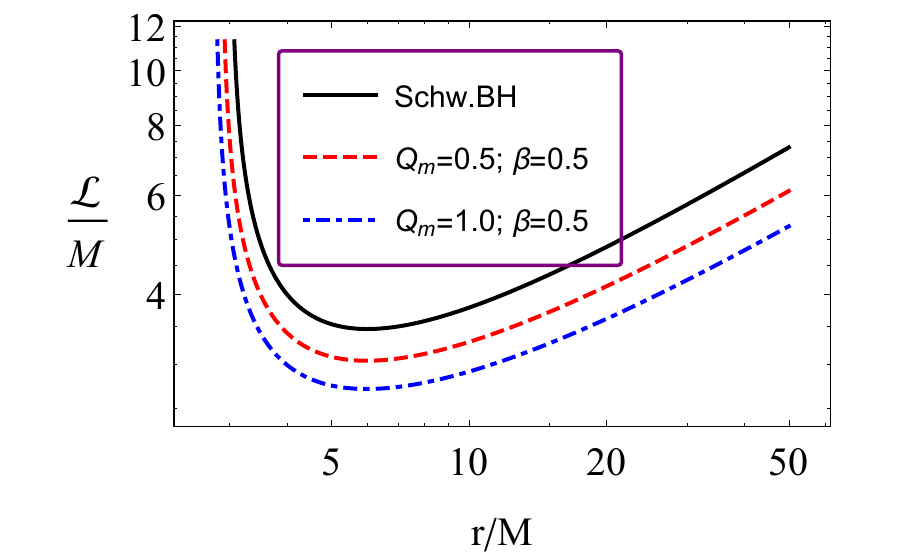}
\includegraphics[width=0.98\linewidth]{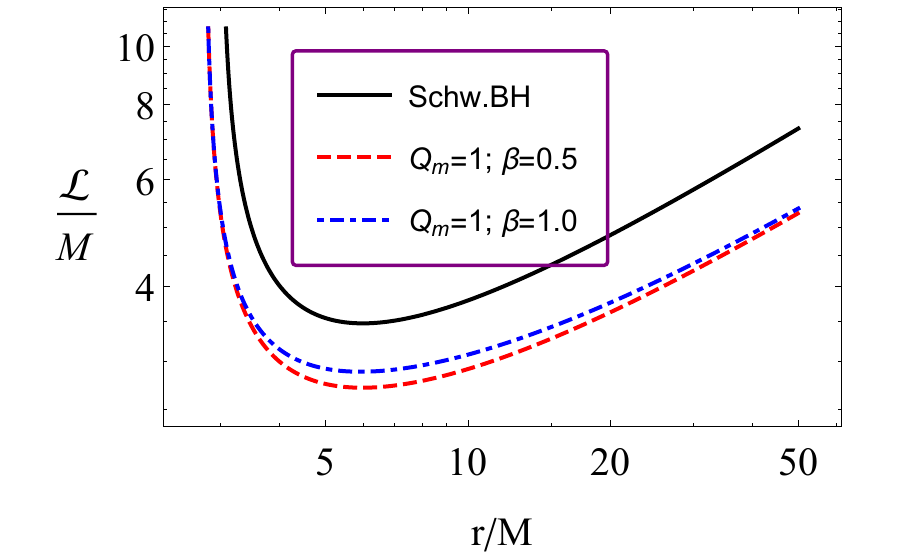} \caption{Radial profile of the specific angular momentum in circular orbits of a magnetic dipole for different values of the magnetic charge $Q_m$ and of the parameter $\beta$. \label{LLgeneric}}\end{figure}

Fig.~\ref{LLgeneric} shows the radial profile of the specific angular momentum of a magnetic dipole around a magnetically charged stringy black hole. One can see from the figure that if we increase the magnetic charge of stringy black hole (the parameter $\beta$ for the magnetic dipole), the specific angular momentum of the magnetic dipole decreases and the inner circular orbit comes closer to the central object, while the parameter $\beta$ does not change the distance of the last circular orbit.

\begin{figure}[ht!]\centering\includegraphics[width=0.98\linewidth]{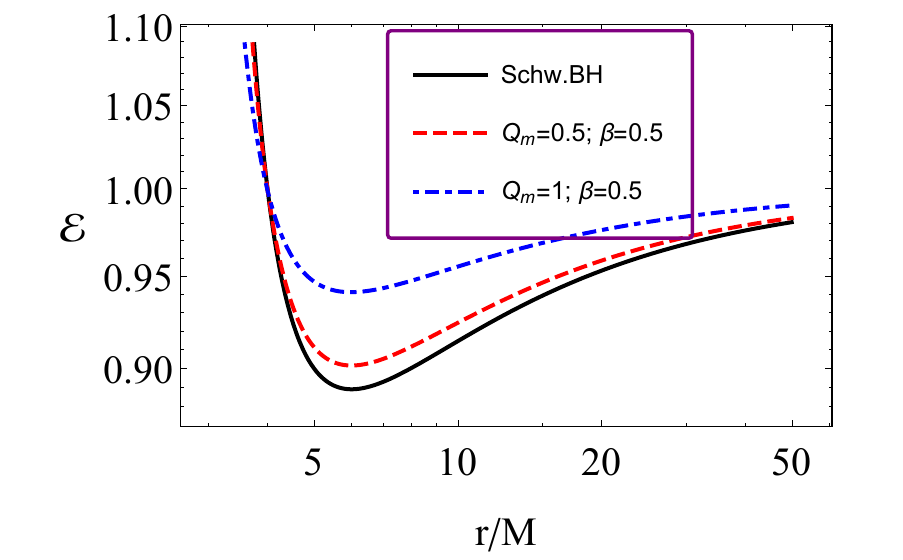}
\includegraphics[width=0.98\linewidth]{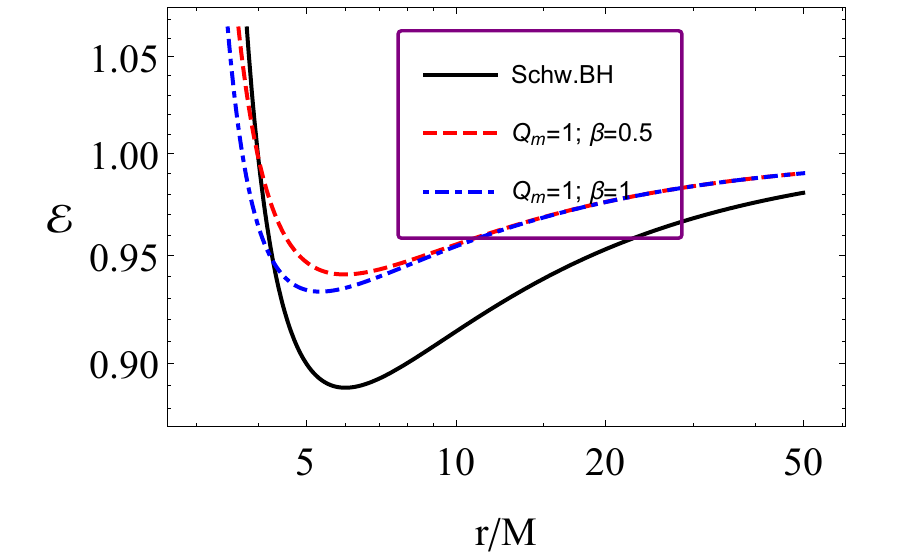} \caption{Radial profile of the specific energy in circular orbits of a magnetic dipole for different values of the magnetic charge $Q_m$ and of the parameter $\beta$. \label{EEgeneric}}\end{figure}

The radial profile of the specific energy of a magnetic dipole for different values of the parameter $\beta$ and of the black hole magnetic charge $Q_m$ is shown in Fig.~\ref{EEgeneric}. One can see from Fig.~\ref{EEgeneric} that the increase of both the black hole magnetic charge $Q_m$ and of the parameter $\beta$ make the specific energy of the magnetic dipole in circular orbits increase. However, the effect of the magnetic charge is stronger than the effect of the parameter $\beta$.

We can get the equation for the ISCO radius taking into account the conditions (\ref{conditions}) for the effective potential (\ref{effpot}) in the following form
\begin{eqnarray}\label{ISCOEQ}
\nonumber
r^5 (r-6 M) \left(2 M^2-Q_m^2\right)+\mathcal{B}^2 \Big[3 r^2 \left(Q_m^2+14 M^2\right)\\
-6 M r \left(3 Q_m^2+10 M^2\right)+32 M^2 Q_m^2-8 M r^3\Big]\geq 0\ . 
\end{eqnarray}
%

 %
\begin{figure}[ht!]\centering\includegraphics[width=0.98\linewidth]{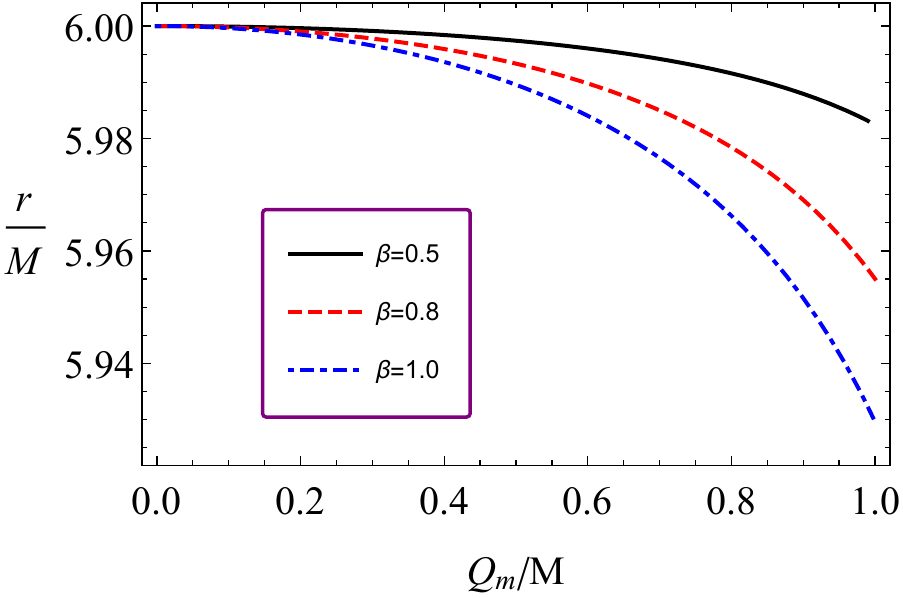}
\includegraphics[width=0.98\linewidth]{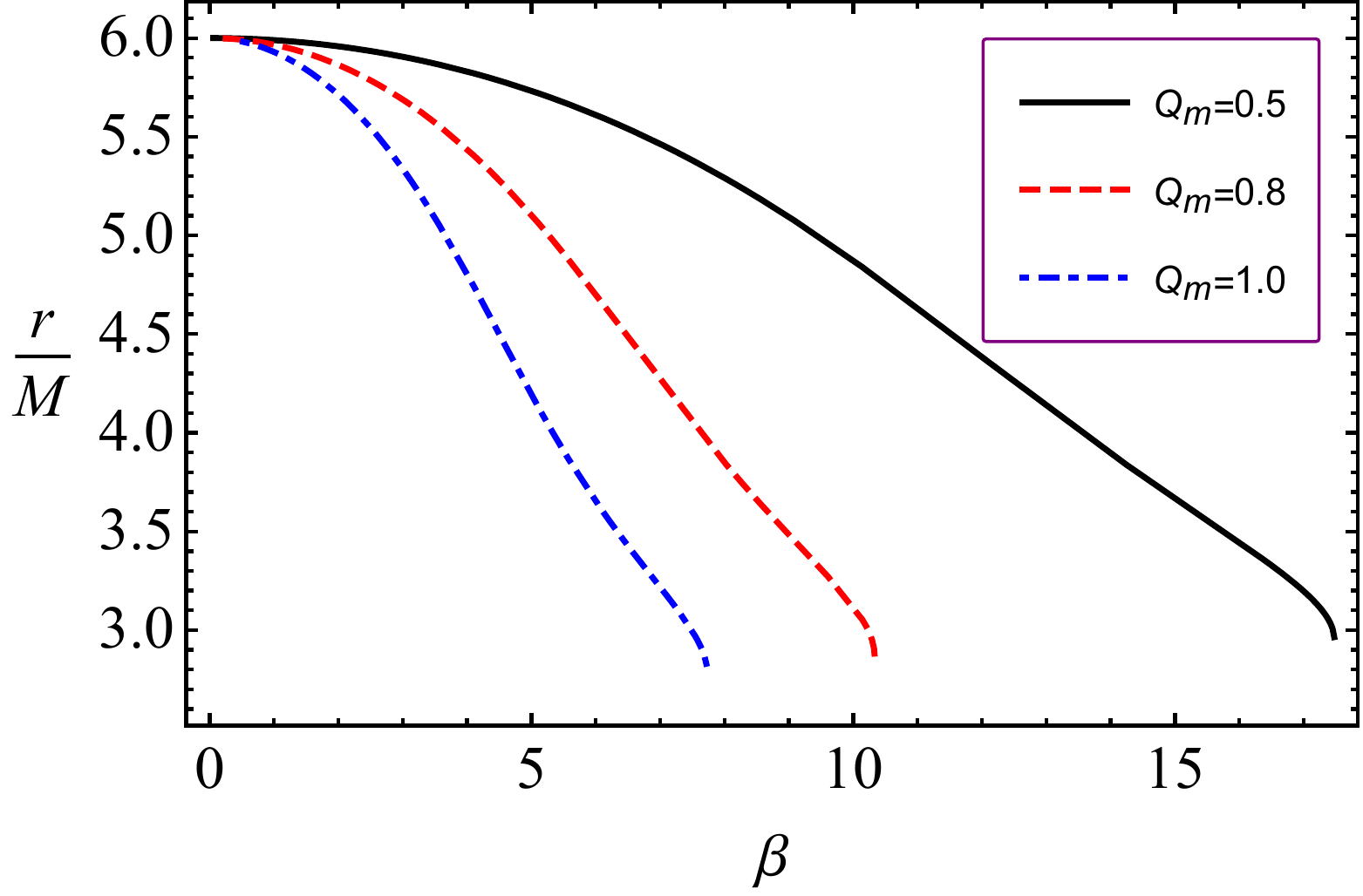} 
\caption{ISCO radius of a magnetic dipole around magnetically charged stringy black hole for different values of the parameter $\beta$ and the black hole magnetic charge. The top panel shows the impact of the parameter $\beta$ and the bottom panel shows the impact of the black hole magnetic charge. \label{ISCOgeneric}}\end{figure}

The numerical solution of Eq.~(\ref{ISCOEQ}) with respect to the radial coordinate is presented in Fig.~\ref{ISCOgeneric}, where we show the ISCO radius as a function of the black hole magnetic charge and of the parameter $\beta$. We can see that the ISCO radius decreases if the magnetic charge parameter $Q_{\rm m}$ increases, and the decreasing rate increases when we increase the parameter $\beta$.  Moreover, an upper value for the parameter $\beta$ exists. For $\beta$ exceeding such an upper value, there are no circular stable orbits for the magnetic dipole. This upper value for the parameter $\beta$ decreases with the increase of the magnetic charge parameter. However, the value of the ISCO radius is the same for all values of the magnetic charge parameter at the upper value of the parameter $\beta$.

\section{Astrophysical applications\label{application}}

In this section, we would like to answer to the following question: can the magnetic charge of a stringy black hole mimic the spin of a rapidly rotating Kerr black hole through its magnetic interactions with orbiting material? To address this question, we study the ISCO radius of magnetic dipoles around $(i)$ a magnetically charged stringy black hole, $(ii)$ a rapidly rotating Kerr black hole, and $(iii)$ a Schwarzschild black hole immersed in an external, asymptotically uniform, magnetic field. We are going to show the cases when the magnetic charge mimics the spin and magnetic interaction parameters providing the same ISCO radius.

We note that we focus on the location of the ISCO because the latter is often the key-quantity in the interpretation of the electromagnetic spectra of black holes~\cite{Bambi17c,Krawczynski18}. The radiation emitted by material at larger radii is not very informative about the spacetime metric: it is normally difficult to determine the orbital radius because relativistic effects are quite similar and at larger radii relativistic effects are weaker. The radiation emitted by material inside the ISCO is normally negligible and difficult to model: there are no stable circular orbits inside the ISCO, so when a particle reaches the ISCO it quickly plunges onto the black hole. In the end, the ISCO is the most sensitive quantity of the spacetime metric and it is relatively easy to measure with electromagnetic observations. When the electromagnetic spectrum of a black hole is dominated by the thermal spectrum of its accretion disk and the Eddington-scaled disk luminosity is between $\sim 5$\% to $\sim 30$\%, the inner edge of the disk is at the ISCO of the spacetime with a good approximation; see~\cite{Steiner10,McClintock14} and reference therein.

In General Relativity, the radial coordinate of the ISCO radius has not a direct physical meaning, as it depends on the coordinate system, but still the value of the ISCO radius is normally a good proxy to compare black hole spacetimes with similar observational properties in the electromagnetic spectrum~\cite{Bambi17c,Krawczynski18}. Thermal spectra of thin accretion disks around black holes are multi-temperature blackbody spectra with a high energy cutoff determined by the inner edge of the disk, so by the ISCO radius~\cite{Kong14}: if we fit the data with a metric in which the ISCO radius is determined by two parameters, we find the typical banana shape in the plot of those parameters and we cannot measure simultaneously the two parameters~\cite{Tripathi:2020qco}. A similar problem is found in the analysis of the reflection spectrum of the disk~\cite{Bambi17b,Cao17}, even if in the presence of high quality data and for an ISCO radius very close to the black hole event horizon it is possible to break such a parameter degeneracy~\cite{Tripathi:2018lhx,Zhang:2019ldz}. Polarimetric measurements are also affected by the same issue~\cite{Krawczynski12}.

The ISCO radius for prograde and retrograde orbits of a test particle around a rotating Kerr black hole can be expressed with the following compact formula~\cite{Bardeen72}
\begin{eqnarray}
r_{\rm isco}= 3 + Z_2 \pm \sqrt{(3- Z_1)(3+ Z_1 +2 Z_2 )} \ ,
\end{eqnarray}
where
\begin{eqnarray} \nonumber
Z_1 &  = & 
1+\left( \sqrt[3]{1+a}+ \sqrt[3]{1-a} \right) 
\sqrt[3]{1-a^2} \ ,
\\ \nonumber
Z_2 & = & \sqrt{3 a^2 + Z_1^2} \ .
\end{eqnarray}

We plan to perform the above-mentioned study of the ISCO analyzing the motion of the magnetar SGR (PSR) J1745--2900 orbiting Sgr A*. We model the magnetar as a test particle with magnetic dipole. 

The magnetar called SGR (PSR) J1745--2900 was discovered in 2013 in the radio band~\cite{Mori2013ApJ}. It is orbiting the supermassive black hole Sgr~A*, whose mass is $M \approx 3.8 \times 10^6M_{\odot}$. From the analysis reported in~\cite{Mori2013ApJ}, we can estimate the value of the parameter $\beta$. The magnetic dipole moment of the magnetar is $\mu \approx 1.6\times 10^{32} \rm G\cdot cm^3$ and its mass is $m \approx 1.5 M_{\odot}$. We thus find
\begin{eqnarray}
 \beta=\frac{\mu_{\rm PSR\, J1745-2900}}{m_{\rm PSR\, J1745-2900} M_{\rm SgrA *}}\approx 10.2\ .
 \end{eqnarray} 

\begin{figure}[ht!]\centering
\includegraphics[width=0.98\linewidth]{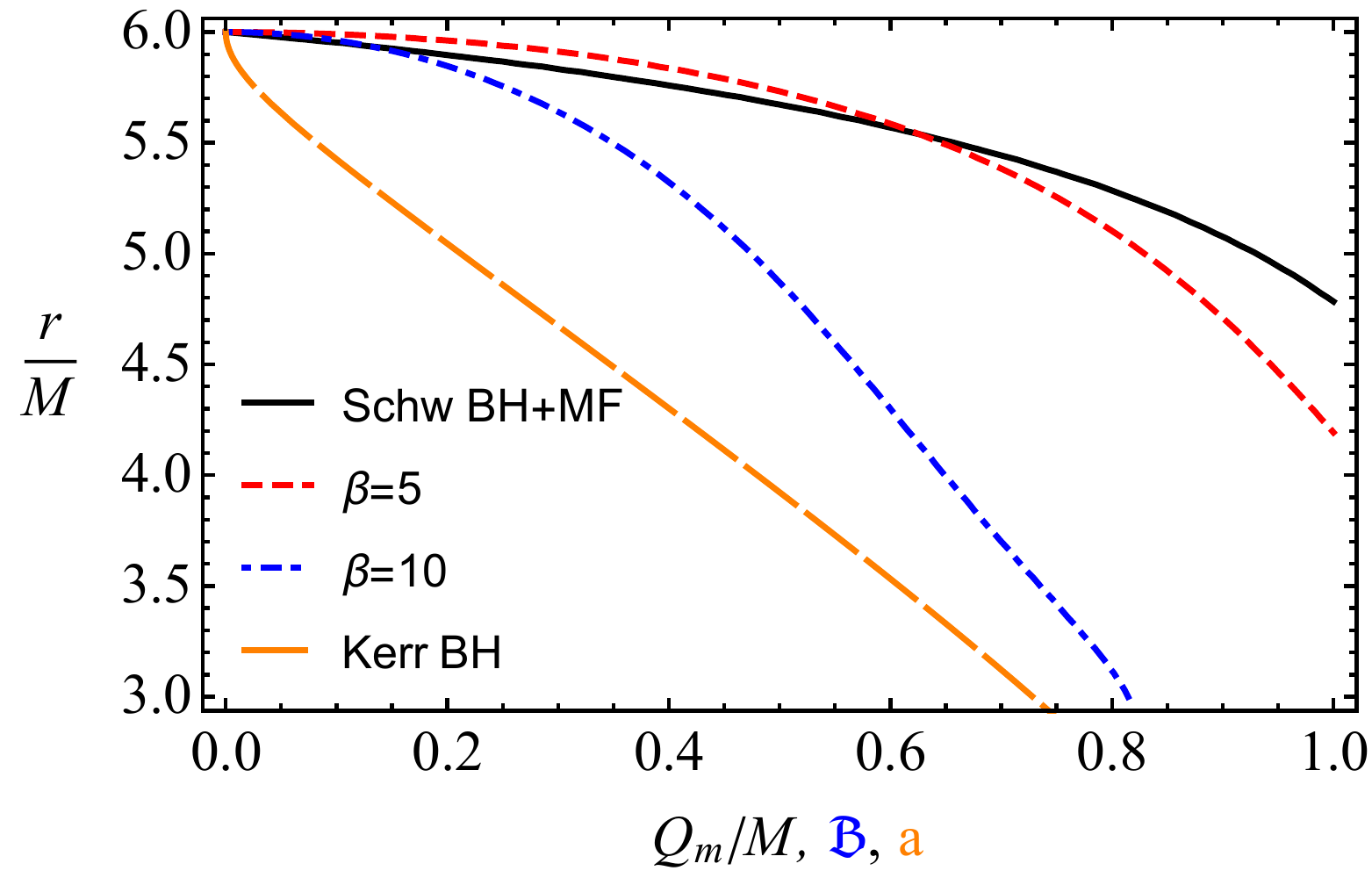}
\caption{ISCO radius of a magnetic dipole orbiting magnetically charged stringy black holes, Kerr black holes, and Schwarzschild black holes immersed in external magnetic fields for different values of the parameter $\beta$. \label{ISCOall}}\end{figure}

Fig.~\ref{ISCOall} presents the ISCO radius of a magnetic dipole around rotating Kerr black holes, Schwarzschild black holes immersed in external magnetic fields, and magnetically charged stringy black holes as a function of their proper parameters $a/M \in (0, 1)$, $Q_{\rm m} \in (0,1)$, and ${\cal B} \in (0, 1)$, respectively. We can see that the effect of the magnetic charge $Q_{\rm m}$ is stronger than the effect of the external magnetic field and it become stronger if we increase the parameter $\beta$.

\subsection{Magnetically charged stringy black holes versus rotating Kerr black holes}

First, we consider the motion of magnetic dipoles and non-magnetized particles around magnetically charged stringy black holes and rotating Kerr black holes, respectively. We will show how the magnetic charge of a stringy black hole can mimic the spin of a Kerr black hole providing the same ISCO radius.

\begin{figure}[h!]\centering\includegraphics[width=0.98\linewidth]{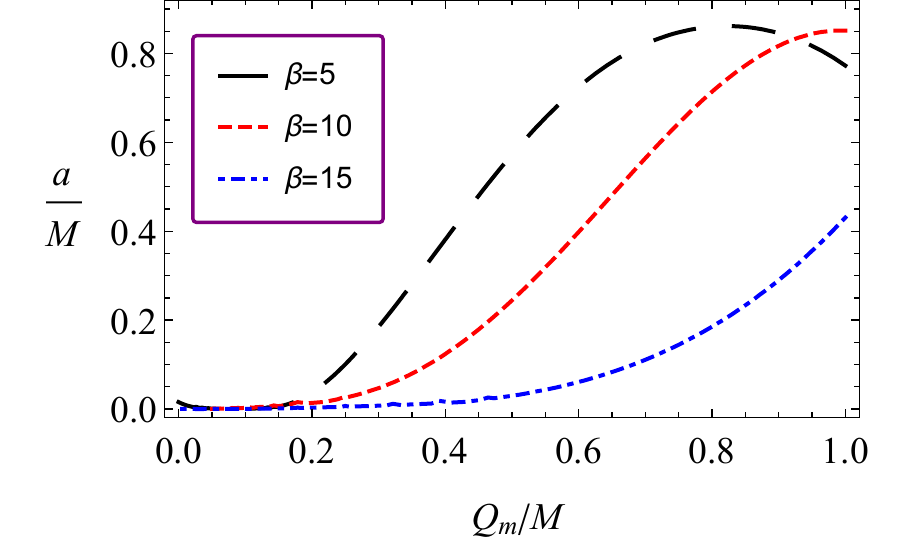}\caption{Relations between the spin of a rotating Kerr black hole and the magnetic charge of a stringy black hole providing the same ISCO radius for different values of the parameter $\beta$. \label{avsq}}\end{figure}

Fig.~\ref{avsq} illustrates the relation between the rotation parameter of a Kerr black hole and the magnetic charge parameter of a stringy black hole that provide the same value of the ISCO radius. We can see that the magnetic charge of the string black hole can mimic the spin of a Kerr black hole up to about 0.5~$M$ for the magnetic dipole with the coupling parameter $\beta=5$. Such an upper value of the spin parameter increases if we increase the value of the parameter $\beta$, but cannot exceed the value $a_* \approx 0.85$. 

\subsection{Magnetically charged stringy black holes versus Schwarzschild black holes in magnetic fields}

Let us now analyze the role of external magnetic fields on the motion of magnetic dipoles and how the magnetic charge of a stringy black hole can mimic the magnetic interaction between an external magnetic field and a magnetic dipole. The topic of magnetic dipole motion around Schwarzschild black holes immersed in external, asymptotically uniform, magnetic fields was first studied by de Felice in~\cite{deFelice}. In Ref.~\cite{Haydarov20b}, we  extended that study to the motion of magnetic dipoles and we showed how the magnetic interaction can mimic a non-rotating black hole in modified gravity (MOG), which is a scalar-tensor-vector theory proposed in~\cite{Moffat06}. The magnetic dipoles dynamics around black holes in conformal gravity~\cite{Haydarov20} and 4-D Einstein Gauss-Bonnet gravity~\cite{Rayimbaev4DEGB2020} has also shown the degeneracy of the magnetic interaction parameters with the spin of Kerr black holes, where a magnetized neutron star was treated as a magnetic dipole with the magnetic coupling parameter $$\mathfrak{B}=\frac{2\mu B_0}{m} =\frac{B_{\rm NS }R^3_{\rm NS}B_{\rm ext}}{m_{\rm NS}}$$ orbiting around a supermassive black hole. Plugging in typical values for a neutron star and an external magnetic field, we find
\begin{eqnarray}
\mathfrak{B}= 0.0044\left(\frac{B_{\rm NS }}{10^{12} \rm G}\right)\left(\frac{R_{\rm NS}}{10^6 \rm cm}\right)^3\left(\frac{B_{\rm ext}}{10\rm G}\right)\left(\frac{m_{\rm NS}}{M_{\odot}}\right)^{-1}.
\end{eqnarray}

We can estimate the value of the interaction parameter for the case of the magnetar SGR (PSR) J1745--2900 orbiting around Sgr A* and we obtain
\begin{equation}
    \mathfrak{B}_{\rm PSR J1745-2900}\simeq 0.716 \left(\frac{B_{\rm ext}}{10\rm G}\right)\ .
\end{equation}

In previous studies \cite{Haydarov20,Haydarov20b,Rayimbaev4DEGB2020}, it was shown that a magnetic dipole with a magnetic interaction parameter $\mathfrak{B}\geq 1$ cannot be in a stable circular orbits due to the destructive nature of magnetic fields. This implies that one can estimate an upper limit for the value of external magnetic fields through such a condition and then predict that the orbit of the magnetar SGR (PSR) J1745--2900 around Sgr A* is stable or not by using $\mathfrak{B} < 1$. Simple calculations show that one can expect that circular orbits of the magnetar are stable only if the external magnetic field in the environment of SrgA* is $B_{\rm ext}\lesssim 14 \rm G $.  This indicates that a magnetar with a surface magnetic field ($B_{\rm surf}$) of the order of $B_{\rm surf}>10^{14} \rm G$ cannot be in a stable orbit in the environment of a supermassive black hole when the external magnetic field is more than about $10 \rm G$. Since the expected magnetic field near Sgr A* is around $100 \rm G$, the magnetic coupling parameter for the magnetar (SGR) PSR J1745-2900 is $\mathfrak{B}\simeq 7.16$ and one can expect to observe pulsars with a surface magnetic field less than $10^{12}$ G. Non-observability of radio pulsars and magnetars in the central part of our Galaxy in vicinity of SgrA* can be caused by either their nonexistence in the region close to ISCO or scattering of radio signals broadening them, which leads to pulsar's signal disappearance. The detailed analysis performed here shows that the interaction of an ambient magnetic field with the magnetar's (pulsar's) magnetic moment is so strong that its orbit in close vicinity of SgrA* would become very unstable and would be unlikely to find a magnetar there. The only opportunity is to look for radio pulsars with low surface magnetic field in that area.

Now we are back to the question if the magnetic charge of the stringy black hole can mimic magnetic field effects providing the same ISCO radius. One can compare the ISCO radius of the two cases of the magnetic dipole motion around a Schwarzschild black hole immersed in an external magnetic field and a magnetically charged stringy black hole following the results of Ref.~\cite{deFelice} and Eq.(\ref{ISCOEQ}).

As the next step, we will focus on the possible degeneracy due to effect of the magnetic coupling parameter and the magnetic charge of the stringy black hole.

\begin{figure}[ht!]\centering\includegraphics[width=0.98\linewidth]{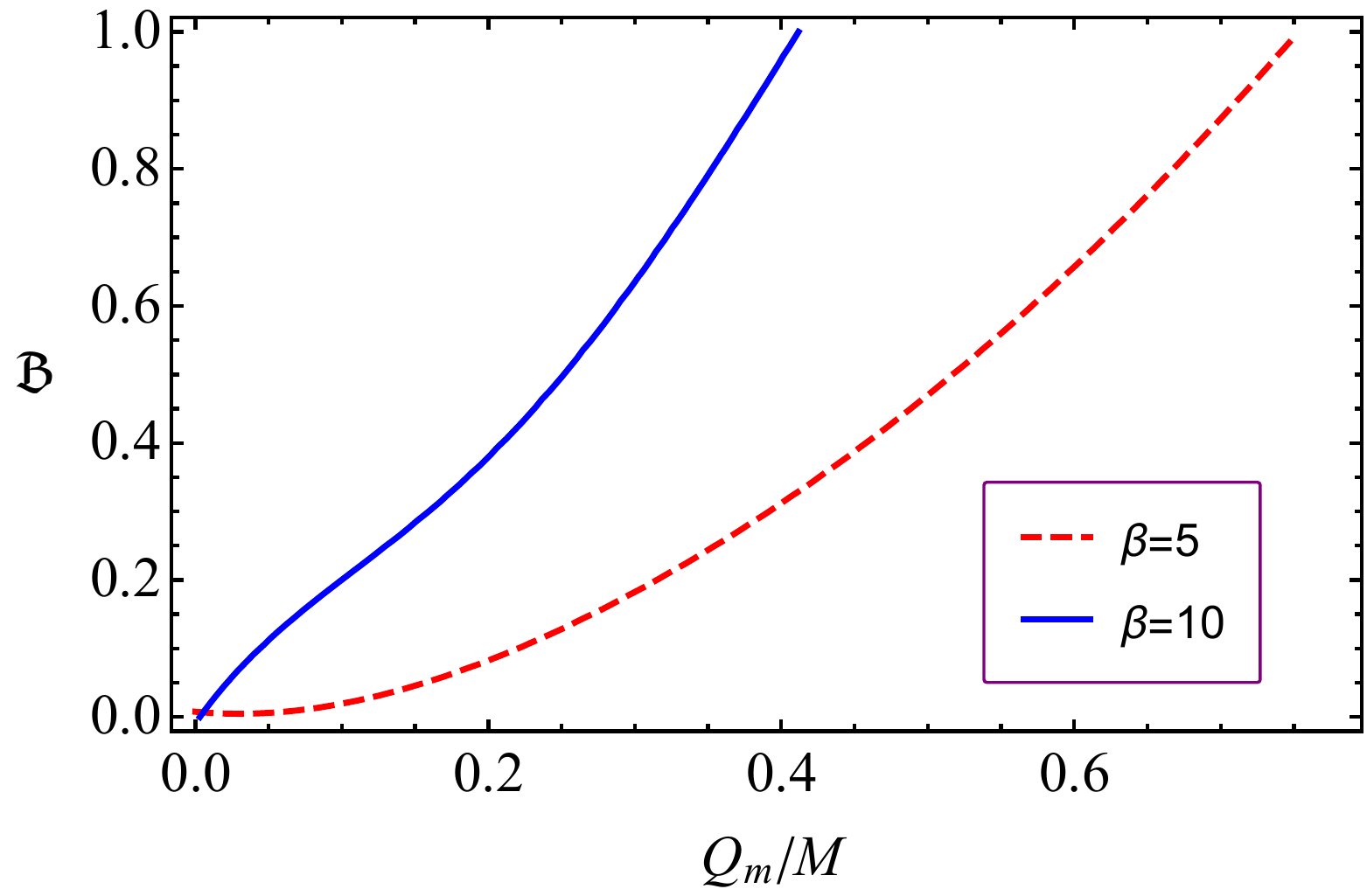}\caption{Relation between the parameters of magnetic coupling and magnetic charge for the same ISCO radius for different values of the parameter $\beta$. \label{betavsq}}\end{figure}

Fig.~\ref{betavsq} demonstrates the relation between the magnetic coupling parameter and magnetic charge parameter of a stringy black hole. One may see that the magnetic coupling parameter may mimic the magnetic charge of a black hole up to $Q_{\rm m}/M=0.7532$ for the magnetic dipole with the parameter $\beta=5$ while it mimics up to $Q_{\rm m}/M=0.4118$ for the particle with the parameter $\beta=10$. This implies that when we apply such a result to the case of the magnetar SGR (PSR) J1745--2900 orbiting around Sgr A* it is impossible to distinguish the effects of an external magnetic field with $B \leq 14 \rm G$ and a magnetic charge of a stringy supermassive black hole with the magnetic charge $Q_{\rm m}/M \leq 0.4118$. We hope that the estimation for a realistic case may help to perform studies of magnetic dipoles such as radio pulsars which can be observed as recycled pulsars and magnetars motion around the supermassive black hole SrgA* in the near future when such observations will be possible.

\section{Conclusions}\label{conclusion}

\begin{itemize}
\item First, we have studied the motion of an electrically charged particle around an electrically charged stringy black hole. If the black hole electric charge increases, the ISCO radius of the charged particle decreases. For the maximal value of the black hole electric charge $Q_{\rm ext} = \sqrt{2}M$, we have found that there is a critical value for the particle electric charge $\abs{q_{\rm ext}}=Q_{\rm ext}/2$ such that if the particle electric charge exceeds this value the ISCO radius becomes infinitely large; that is, particles with an electric charge exceeding $q_{\rm ext}$ have no stable circular orbits around the stringy black hole. We have also shown that an electrically charged stringy black hole can assume the same ISCO radius as a Kerr black hole, suggesting that a similar object may mimic well Kerr black holes of any spin.

\item Second, we have studied the motion of magnetic monopoles in the spacetime of a magnetically charged stringy black hole. The magnetically charged black hole solution recovers the Schwarzschild one in the case of vanishing magnetic charge $Q_m$. We have found that the event horizon and the ISCO radius are not affected by the value of $Q_m$ and they have thus the same value as in the Schwarzschild spacetime. This leads to important implications from the observational point of view and suggests that it would challenging for a faraway observer to distinguish a static and spherically symmetric Schwarzschild black hole from a magnetically charged stringy black hole. The ISCO radius changes only for particles with a non-vanishing magnetic charge. The ISCO radius increases if the particle electric charge $g$ is positive and increases. The ISCO radius decreases if the particle electric charge $g$ is negative and its absolute value increases. The ISCO radius cannot get arbitrarily close to the event horizon in the case of negative values of the particle magnetic charge $g$.

\item Last, we have studied the dynamics of magnetic dipoles around magnetically charged stringy black holes in the weak interaction limit introducing a new parameter $\beta$, describing the interaction between the particle magnetic dipole and the central object. From the study of the ISCO radius of magnetic dipoles, we found that the existence of an ISCO is determined by the values of the parameter $\beta$ and of the black hole magnetic charge. If the parameter $\beta$ exceeds a critical value, set by the black hole magnetic charge, there are no stable circular orbits for the magnetic dipole due to the increase of destructive Lorentz forces. Finally, we have investigated how the magnetic charge of a stringy black hole can mimic the spin of a Kerr black hole and the interaction between magnetic dipoles and external magnetic fields providing the same ISCO radius. Our results show that the magnetic charge of a stringy black hole can mimic the spin parameter of a Kerr black hole up to $a_* \sim 0.85$, while the magnetic interaction parameter can mimic the magnetic charge effects up to $Q_{\rm m}/M=0.7532$ for a magnetic dipole when $\beta=5$. We applied these findings to the magnetar SGR (PSR) J1745 orbiting around the SMBH Srg~A* and we argued that the latter may be a stringy black hole with magnetic charge up to $Q_{\rm m}/M=0.4118$. 
\end{itemize}

\section*{Acknowledgement}

This research is supported by Grants No. VA-FA-F-2-008, No.MRB-AN-2019-29 of the Uzbekistan Ministry for Innovative Development. 
JR, AA and BA thank Silesian University in Opava for the hospitality during their visit. 
B.N. acknowledges support from the China Scholarship Council (CSC), grant No.~2018DFH009013.
The work of B.N. and C.B. was supported by the Innovation Program of the Shanghai Municipal Education Commission, Grant No.~2019-01-07-00-07-E00035, and the National Natural Science Foundation of China (NSFC), Grant No.~11973019.
\bibliographystyle{apsrev4-1}
\bibliography{gravreferences}

\end{document}